\documentclass[12pt,journal,draftclsnofoot,onecolumn]{IEEEtran}
\usepackage[T1]{fontenc}

\usepackage{amsmath}
\usepackage[cmintegrals]{newtxmath}
\usepackage{bm}

\newtheorem{Proposition}{\it Proposition}[section]

\makeatletter
\newcommand{\Rmnum}[1]{\expandafter\@slowromancap\romannumeral #1@}
\makeatother
\usepackage{verbatim}
\usepackage{graphicx}
\usepackage{multicol}
\usepackage{setspace}
\usepackage{subfigure}
\usepackage{amsmath}
\usepackage{epstopdf}
\usepackage{fancyhdr} 
\usepackage{enumerate}

\usepackage{amssymb}
\usepackage{stfloats}
\usepackage{multirow}
\usepackage{threeparttable}
\usepackage[justification=centering]{caption}
\usepackage{makecell}
\usepackage{caption}
\usepackage{cases}
\usepackage{algorithm}
\usepackage{algpseudocode}
\usepackage{graphics}
\usepackage{subfig}
\usepackage{epsfig}
\usepackage{color}
\usepackage{colortbl}
\usepackage{cite}

\begin{document}
%
\title{Universal Performance Bounds for Joint Self-Interference Cancellation and Data Detection in Full-Duplex Communications}
%
%
%

\author{Meng He,~\IEEEmembership{Student Member,~IEEE,}
	and Chuan Huang,~\IEEEmembership{Member,~IEEE}
\thanks{Part of this paper was presented in IEEE Globecom 2019\cite{9013150}.
	
M. He and C. Huang are with the Future Network of Intelligence Institute and the School of Science and Engineering, The Chinese University of Hong Kong, Shenzhen, China, 518172 (email: menghe@link.cuhk.edu.cn and huangchuan@cuhk.edu.cn).}}

\maketitle

\begin{abstract}
	
This paper studies the joint digital self-interference (SI) cancellation and data detection in an orthogonal-frequency-division-multiplexing (OFDM) full-duplex (FD) system, considering the effect of phase noise introduced by the oscillators at both the local transmitter and receiver. In particular, an universal iterative two-stage joint SI cancellation and data detection framework is considered and its performance bound independent of any specific estimation and detection methods is derived. First, the channel and phase noise estimation mean square error (MSE) lower bounds in each iteration are derived by analyzing the Fisher information of the received signal. Then, by substituting the derived MSE lower bound into the SINR expression, which is related to the channel and phase noise estimation MSE, the SINR upper bound in each iteration is computed. Finally, by exploiting the SINR upper bound and the transition information of the detection errors between two adjacent iterations, the universal bit error rate (BER) lower bound for data detection is derived.

\end{abstract}


%
\IEEEpeerreviewmaketitle

\section{Introduction}
Full-Duplex (FD) technology is a promising mechanism that can significantly improve the spectral efficiency of future wireless communication systems compared to the conventional half-duplex technologies \cite{6832592, 7736037, 7051286}, since it allows the wireless transceiver to simultaneously transmit and receive signals over the same frequency band \cite{6319352}. 
However, simultaneous transmission and reception at one FD transceiver introduces extremely strong self-interference (SI) from the local transmitter to the local receiver\cite{6799314, alves2020full}.
To make the FD transmission practical and feasible with the potential advantages on spectral efficiency, the FD system should efficiently suppress the SI at the receiver to the noise floor \cite{6832464}.

Recently, bunches of literature \cite{6190376, 6702851, 6656015, 123456, 7906500, 5757799, 6353396, 6810483, 6523998, 7815419} have investigated the SI cancellation in the FD systems, and according to \cite{7051286}, the SI cancellation methods are categorized into two types: passive suppression and active cancellation. 
Passive suppression techniques\cite{6190376, 6702851, 6656015} eliminated the SI signal in the propagation domain before it is received by the local receiver. More specifically, these techniques tried to isolate the local transmitter antennas from the local receiver antennas \cite{6702851} by adopting the directional antennas \cite{7736037,123456,6353396}, absorptive shielding \cite{6702851}, and cross-polarization \cite{7906500}.
In contrast, active cancellation mechanisms\cite{5757799, 7906500,  6353396, 6810483, 6523998} mitigated the SI signal in the digital and analog domains, where a reconstructed SI cancellation signal is subtracted from the received signal to cancel the strong SI. The reconstruction of the SI signal is based on the information transportation between the local transmitter and receiver, and the estimation of the unknown SI channel \cite{7815419}.

However, the SI cannot be completely mitigated in practical systems. Due to the mismatch between the SI signal and the reconstructed one, the residual SI always exists\cite{6151328} and limits the performance of various FD systems \cite{ 6854102, 6488952, 7273946}. In particular, phase noise introduced by the oscillator defects was identified to be one of the main causes of the cancellation mismatch\cite{6523998, 6489379, 7051286}. In OFDM systems, phase noise introduces both the common phase error (CPE) and the inter-carrier interference (ICI). To analyze and mitigate the phase noise in the OFDM FD systems, phase noise estimation and suppression mechanisms were proposed in\cite{6799314, 6523998, 6810483, 7815419, 6937196, 7145898, 7343306}. To analyze the oscillator phase noise effects on the SI cancellation capability, the authors in \cite{6799314} derived the closed-form expression for the power of the residual SI, considering both the two cases with two independent oscillators and one common oscillators at one transceiver. Considering the SI channel estimation in the presence of phase noise, the authors in \cite{7145898} adopted the expectation maximization (EM) scheme to jointly estimate the SI channel and CPE, and the simulation results indicated that better SI suppression performance can be achieved by considering the phase noise effects. They also predicted that an ICI mitigation scheme may further increase the SI cancellation capacity. Considering the CPE estimation, the authors in \cite{7815419} analytically derived the closed-form expression for the digital SI cancellation capability in terms of the power of the CPE, the interference-to-noise-ratio (INR), and the signal-to-noise-ratio (SNR), and concluded that the residual ICI severely limits the digital cancellation ability. To suppress the phase noise ICI, the authors in \cite{6937196} proposed one frequency-domain and one time-domain ICI suppression methods. By the theoretical analysis and simulations, they investigated the feasibility of these two techniques in terms of complexity and achievable ICI cancellation gain, and concluded that the SI cancellation ability of the proposed ICI suppression methods is limited by the power of the signal-of-interest (SoI).  

In this work, we consider the joint digital SI cancellation and data detection problem in an OFDM FD two-way system with phase noise introduced by the oscillators at both the transmitter and  the receiver \cite{6937196}.
In particular, we consider a typical scenario that the coherence time of the SI channel is much longer compared with the duration of an OFDM symbol, and that the phase noise and the SoI channel vary much faster than the SI channel.
Thus, a two-stage framework is adopted to estimate the SI channel  in the first stage and detect the desired data in the second stage. In the SI channel estimation stage, one pilot OFDM symbol is transmitted from the remote transmitter to the local receiver, and all the unknown channel coefficients and phase noise are jointly estimated.
In the data transmission stage, the mixed pilot and data OFDM symbol is transmitted to the local receiver, and then the unknown SI phase noise, SoI channel and the desired data are iteratively estimated and detected.
Next, we investigate the universal estimation and detection performance for the considered two-stage framework.
The MSE lower bounds for the  estimation of SoI channel and SI phase noise  in each iteration are derived by analyzing the Fisher information of the received signal, considering the data detection results from the previous iteration.
Then, by subsisting the derived estimation MSE lower bounds into the effective SINR expression, which in related to the channel and phase noise estimation MSE, the effective SINR upper bound in each iteration is then computed. 
Finally, by exploiting the SINR upper bound and the transition information of the detection errors between two adjacent iterations, the theoretically BER lower bound for the two-stage framework is derived, and is independent of any specific estimation and detection methods.

The remainder of the paper is organized as follows: Section \ref{System Model} introduces the considered FD OFDM two-way system. Section III formulates the SI cancellation and data detection problem, and proposes the two-stage framework for the problem. Then, section \ref{BER Bound in Data Transmission Stage} gives the theoretical performance analysis for  the considered scheme. Next, section \ref{Simulation Analysis} presents the simulation analysis. Finally, section \ref{Conclusion} concludes this paper.

Notations: Boldface small letters, e.g. $ \mathbf{x} $, denote vectors, and boldface capital letters, e.g. $ \mathbf{X} $, denote matrices. $ \mathbf{I}_{N\times M} $ and $ \mathbf{0}_{N\times M} $ represent all one and zero $N\times M$ matrix, respectively. $[\mathbf{X}]_{i,j}$ denotes the $ (i, j) $-th entry of matrix $ \mathbf{X} $. $\mathrm{tr}\{\mathbf{X}\}$ denotes the trace of matrix $\mathbf{X}$. The operator $\mathrm{diag}\{\mathbf{x}\}$ transforms the vector $\mathbf{x}$ into a diagonal matrix. $(\cdot)^*, (\cdot)^T$, $ (\cdot)^H $, and $*$ denote conjugate, transpose, conjugate transpose and convolution operators, respectively. $\mathbb{E}[\cdot]$, $\Re\{\cdot\}$, and $\Im\{\cdot\}$ represent expectation, real and imaginary operators, respectively. $\lfloor\cdot\rfloor$ is the round down operator. $|\cdot|$ denotes the absolute. $\log(\cdot)$ is the base-10 logarithm function. $\Vert{\cdot\Vert}^2$ is the L-2 norm. 
$p(\cdot)$ denotes the probability. $\mathbb{N}^+$ is the positive integer set.

\begin{figure}[htbp]
	\vspace{-10pt}
	\setlength{\abovecaptionskip}{10pt}
	\setlength{\belowcaptionskip}{-5pt}
	\centering
	\includegraphics[width=5.6in]{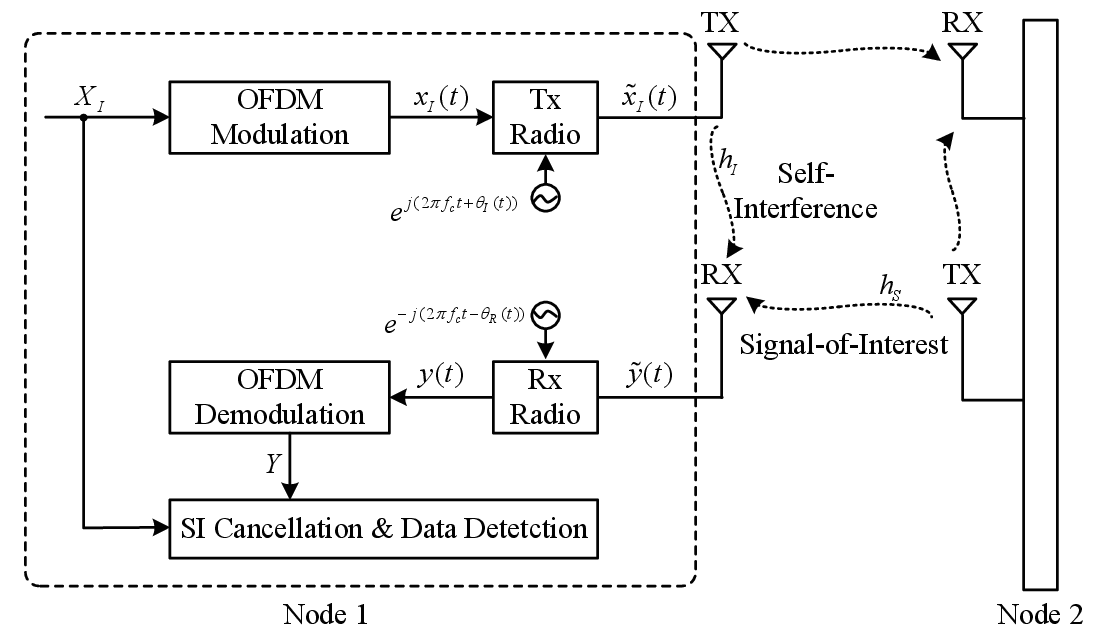}
	\caption{Two-way full-duplex OFDM communication system.}
	\label{fig_1}
\end{figure}
 
\section{System Model}
\label{System Model}
In this paper, a FD OFDM two-way system is considered as shown in Fig. \ref{fig_1}, where two transceivers, labeled as Node 1 and Node 2, respectively, simultaneously transmit and receive signals at the same frequency band. Take Node 1 for example: at its local transmitter, the digital symbols $\{X_I[k]\}_{k=0}^{N-1}$ are first transformed into the time-domain signal $x_I(t)$ by a standard OFDM modulator\cite{7815419}, with $N$ being the total number of subcarriers in one OFDM symbol. Then, at the local transmitter oscillator, $x_I(t)$ is mixed with the oscillator signal $e^{j(2\pi f_c(t)+\phi_I(t))}$ to obtain the radio frequency (RF) signal $\tilde{x}_I(t)$ to be transmitted to Node 2, i.e.,
\begin{equation}
\label{SI RF Signal 1}
\tilde{x}_I(t)=x_I(t)e^{j(2\pi f_ct +\theta_I(t))},
\end{equation}
where $f_c$ is the carrier frequency and $\phi_I(t)$ represents the phase noise at the transmitter oscillator. Notice that this RF signal from Node 1 to Node 2 is also received by the local receiver of Node 1, and becomes a strong SI at the local receiver.   

At the local receiver antenna of Node 1, the SoI from Node 2 is mixed with the SI from the local transmitter of Node 1, and thus the received RF signal $\tilde{y}(t)$ at the receiver antenna is given as
\begin{equation}
\label{Received RF Signal 1}
\tilde{y}(t)=\sum_{l=0}^{L_I-1}h_I(l)\tilde{x}_I(t-\tau_I(l))+
\sum_{l=0}^{L_S-1}h_S(l)\tilde{x}_S(t-\tau_S(l))+w(t),
\end{equation}
where $w(t)$ is the circular symmetric complex Gaussian (CSCG) noise, $\tilde{x}_S(t)$ is the time-domain desired data signal, $\{h_I(l)\}_{l=0}^{L_I-1}$ and $\{h_S(l)\}_{l=0}^{L_S-1}$ represent the time-domain multipath SI and SoI channel impulse, with $L_I$ and $L_S$ being the numbers of the SI and SoI multipath channel taps, respectively. 

At the receiver oscillator, the received RF signal $\tilde{y}(t)$ is mixed with the oscillator signal to be downconverted to the baseband signal $y(t)$, i.e., $y(t)=\tilde{y}(t)e^{-j(2\pi f_c(t)-\phi_R(t))}$, where $\phi_R(t)$ represents the phase noise at the receiver oscillator of Node 1. Then, after the OFDM demodulation, $y(t)$ is restored to the digital frequency-domain symbols $\{Y[k]\}_{k=0}^{N-1}$ \cite{4355336}, i.e,
\begin{equation}
\label{received digital domain signal}
\begin{split}
Y[k]=\sum_{l=0}^{N-1}X_I[l]H_I[l]J_I[k-l]+\sum_{l=0}^{N-1}X_S[l]H_S[l]J_S[k-l]+W[k],
\end{split}
\end{equation}
where $k\in\{0,1,\dots,N-1\}$ is the OFDM subcarrier index, $X_I[k]$ and $X_S[k]$ represent the digital SI and SoI symbols, $H_I[k]$ and $H_S[k]$ denote the frequency-domain SI and SoI channel impulse coefficients, $W[k]$ is the frequency-domain receiver CSCG noise with zero mean and variance $\frac{1}{N}\sigma_w^2$, $J_I[k]$ and $J_S[k]$ denote the frequency-domain SI and SoI phase noises \cite{8403642}, i.e.,
\begin{equation}
\label{DFT coefficients of the phase noise}
\begin{cases}
J_I[k]=\dfrac{1}{N}\sum\limits_{n=0}^{N-1}e^{j(\theta_I(nT_s-t_I)+\theta_R(nT_s))}e^{-j\frac{2\pi kn}{N}},\\
J_S[k]=\dfrac{1}{N}\sum\limits_{n=0}^{N-1}e^{j(\theta_S(nT_s-t_S)+\theta_R(nT_s))}e^{-j\frac{2\pi kn}{N}}, 	
\end{cases}
\end{equation} 
$\theta_S(t)$ represents the phase noise at the transmitter oscillator of Node 2, $T_s$ is the digital sampling time, $t_I$ is the SI transmission delay, and $t_S$ is the SoI transmission delay. 
Compared with the case without the oscillator phase noise, i.e., $Y[k] = H_I[k]X_I[k] + H_S[k]X_S[k] +W[k]$, it is observed from equation \eqref{received digital domain signal} that phase noise destroys the orthogonality of the OFDM subcarriers and introduces the intercarrier interference (ICI). Thus, the existence of oscillator phase noise causes the signal constellation rotation and increases the noise floor, which degrades the SI cancellation performance \cite{7815419}.

Using the matrix notations, the received signal is rewritten as
\begin{equation}
\label{matrix form of receive signal 1}
\mathbf{y}=\mathbf{J_I}\mathbf{H_I}\mathbf{x_I}+\mathbf{J_S}\mathbf{H_S}\mathbf{x_S}+\mathbf{w},
\end{equation}
where $\mathbf{y}\!=\![Y[0],\cdots,Y[N-1]]^T$ is the received signal, $\mathbf{x_I}\!=\![X_I[0],\cdots,X_I[N-1]]^T$ denotes the SI symbol, $\mathbf{x_S}\!=\![X_S[0],\cdots,X_S[N-1]]^T$ is the desired data, $\mathbf{H_I}\!=\!\mathrm{diag}\{H_I[0],H_I[1],\dots,H_I[N-1]\}$ and $\mathbf{H_S}\!=\!\mathrm{diag}\{H_S[0],H_S[1],\dots,H_S[N-1]\}$ are the SI and SoI channel matrices, respectively, and $\mathbf{J_I}$ and $\mathbf{J_S}$ represent the SI and SoI phase noise matrices with their entries being $[\mathbf{J_I}]_{m, n}=J_I[m-n],\,[\mathbf{J_S}]_{m, n}=J_S[m-n]$, $m, n=0,1,\cdots,N-1$, and $\mathbf{w}=[W[0],W[1],\cdots,W[N-1]]^T$ is the CSCG noise vector, $\mathbf{w}\sim\mathcal{N}(\mathbf{0}_{N\times1},\frac{1}{N}\sigma_{w}^2\mathbf{I}_{N\times N})$.

\section{Two-Stage Scheme}
\label{Problem Formulation and Two-Stage Scheme}

In this section, a two-stage joint SI cancellation and data detection framework is adopted to estimate the SI channel in the first stage and detect the desired data in the second stage. According to Fig. \ref{fig_1} and Eq. \eqref{matrix form of receive signal 1}, the SI symbol $\mathbf{x_I}$ at the receiver of Node 1 is known, while the desired data $\mathbf{x_S}$ from Node 2, the channels $\mathbf{H_I}$ and $\mathbf{H_S}$, and the phase noises $\mathbf{J_I}$ and $\mathbf{J_S}$ are all unknown.

\begin{figure}[htbp]
	\vspace{-10pt}
	\setlength{\abovecaptionskip}{-5pt}
	\setlength{\belowcaptionskip}{-5pt}
	\centering
	\includegraphics[width=4in]{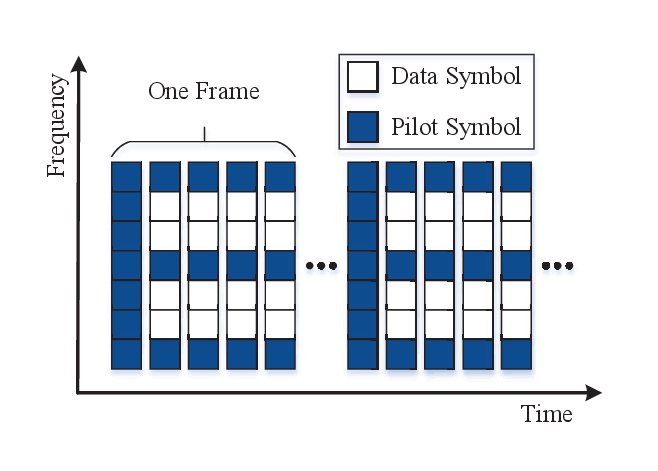}
	\caption{Two-stage joint SI cancellation and data detection for OFDM systems.}
	\label{fig_2}
\end{figure}

Generally, the SI channel is slowly varying across multiple OFDM symbols \cite{7518612}. 
On the other hand, phase noise has a much smaller coherence time compared with that of the SI channel and thus should be repeatedly estimated in each OFDM symbol. 
Accordingly, we can group multiple OFDM symbols as one frame. 
Due to the slowly-varying nature of the SI channel, we only need to estimate its channel state information (CSI) in the first symbol of one frame and use this estimation in the rest symbols of this frame.
Based on the above analysis, the following two-stage framework shown in Fig. \ref{fig_2} is considered to solve the joint SI cancellation and data detection problem for the FD OFDM systems.
\subsubsection{SI Channel Estimation Stage}
The first OFDM symbol of each frame is utilized for the SI channel estimation. 
In this stage, Node 2 transmits a known pilot OFDM symbol to Node 1. 
By exploiting both the pilot symbol from Node 2 and the known SI symbol from the local transmitter,  all unknown channel coefficients and phase noises are jointly estimated.

\subsubsection{Data Transmission Stage}	
The rest OFDM symbols are utilized for the data transmissions. Each OFDM symbol in this stage from Node 2 has both the known pilots and unknown data allocated to different subcarriers. 
Here, by exploiting the known pilots and the SI channel estimations obtained from the SI channel estimation stage, SI phase noise is estimated and the desired data is detected.

\subsection{SI Channel Estimation}
\label{SI Channel Estimation Stage}

In the SI channel estimation stage, Node 2 transmits one pilot OFDM symbol to Node 1 to estimate the SI channel.
At the receiver of Node 1, with the known pilot symbol $\mathbf{x_S}$ from Node 2 and the SI symbol $\mathbf{x_I}$ from the transmitter of Node 1, we are going to solve a conventional joint channel and phase noise estimation problem in this stage,
which has been well studied in \cite{4355336,lin2006joint,lee2017channel,wang2017effective}.
Thus, we do not consider a specific estimation algorithm for this stage and concentrate on the approximation of the channel and phase noise models.

In one OFDM symbol, there are totally $N$ subcarriers, and there are $4N$ unknown parameters to be estimated, i.e., $\{H_I[n]\}_{n=0}^{N-1}$, $\{H_S[n]\}_{n=0}^{N-1}$, $\{J_I[n]\}_{n=0}^{N-1}$, and $\{J_S[n]\}_{n=0}^{N-1}$. However, we can send at most $N$ pilots in the frequency-domain, which is not sufficient to generate a good estimation for these 4$N$ parameters\cite{4355336}. To overcome this difficulty, we obtain the simplified channel and phase noise models with much fewer parameters.

\subsubsection{SI Channel Approximation}
The frequency-domain SI channel $\mathbf{H_I}$ is approximated as $\mathbf{H_I}=\mathrm{diag}\{\mathbf{F_Ih_I}\}$, where $\mathbf{h_I}=[h_I(0),h_I(1),\dots,h_I(L_I-1)]^T$ is the time-domain SI channel impulse, $L_I\ (L_I\ll N)$ is the number of the multipath SI channel taps, and $\mathbf{F_I}$ is a $N \times L_I$ discrete Fourier transform (DFT) matrix 
with each entry being
$[\mathbf{F_I}]_{n, l}=\frac{1}{\sqrt{N}}e^{-j2\pi nl/N}, 0\leq n\leq N-1, 0\leq l\leq L_I-1$.

\subsubsection{SI Phase Noise Approximation}
According to \eqref{DFT coefficients of the phase noise} and \eqref{matrix form of receive signal 1}, the SI phase noise matrix $\mathbf{J_I}$ is determined by vector $\mathbf{j_I}=[J_I[0],J_I[1],\dots,J_I[N-1]]^T$. 
Then, $\mathbf{j_I}$ can be approximated as $\mathbf{j_I}\approx \mathbf{S_I j_I'}$, where $\mathbf{j_I'}=[J_I[0],\dots,J_I[K-1]]^T$ is a subvector of $\mathbf{j_I}$ $, K\leq N-L_S-L_I$ is a constant factor, and $\mathbf{S_I}$ is the phase noise approximation matrix with each entry being
\begin{equation}
\label{approximation model of SI phase noise 1}
\begin{split}
&[\mathbf{S_I}]_{n, k}=
\begin{cases}
1, &  n=k\leq K,\\
0, &  \mathrm{others}, 
\end{cases}\\
&0\leq n\leq N-1, 0\leq k\leq K-1.	
\end{split}
\end{equation}
\subsubsection{SoI Channel and Phase Noise Approximation}
Generally, the ICI of the SoI phase noise has much smaller power than the noise floor at the receiver \cite{6937196}. 
Thus, different from the SI phase noise approximation considering both the CPE and ICI, the frequency-domain SoI channel mixed with the SoI phase noise can be approximated as a CPE-rotated channel $\mathbf{H_D}$\cite{7815419}, i.e.,
\begin{equation}
\label{approximate model of SoI channel and SoI phase noise 1}
\mathbf{J_SH_S}\approx\mathbf{H_D}=\mathrm{diag}\{\mathbf{F_Sh_D}\},
\end{equation}
where $\mathbf{H_D}=J_S[0]\cdot\mathrm{diag}\{[H_S[0],H_S[1]\dots,H_S[N-1]]\}$ is the frequency CPE-rotated SoI channel matrix, $\mathbf{h_D}=J_S[0]\cdot[h_S(0),\dots,h_S(L_S-1)]$ is the time-domain CPE-rotated SoI channel impulse, $L_S\ (L_S\ll N)$ is the number of the SoI channel taps, and $\mathbf{F_S}$ is an $N \times L_S$ DFT matrix.

Based on the approximation models above, the total number of unknown parameters in this stage is reduced to $L_I+L_S+K$. After the SI channel estimation stage, the receiver obtains the SI channel estimation $\mathbf{\hat{h}_I}$, which is used in the next data transmission stage.

\subsection{Data Transmission Stage}
\label{Data Transmission Stage}
\begin{figure}[htbp]
	\vspace{-10pt}
	\setlength{\abovecaptionskip}{10pt}
	\setlength{\belowcaptionskip}{-5pt}
	\centering
	\includegraphics[width=12cm]{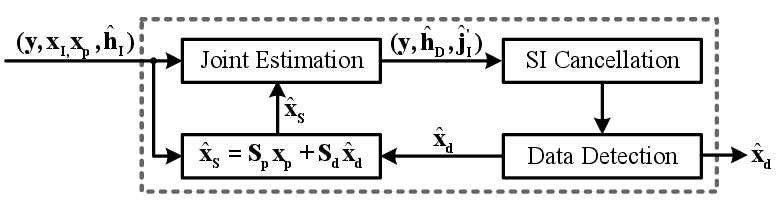}
	\caption{Data transmission stage.}
	\label{Data_Transmission_Stage_1}
\end{figure}
In the rest time of one two-stage frame, i.e. the data transmission stage, Node 2 transmits the mixed pilot and data OFDM symbols to Node 1. According to Fig. \ref{fig_2}, the symbol $\mathbf{x_S}$ from Node 2 in this stage is expressed as
\begin{equation}
\label{x_S,X_p,X_d}
\mathbf{x_S}=\mathbf{S_px_{p}}+\mathbf{S_dx_{d}},
\end{equation}
where $\mathbf{x_p}=[X_S[i_0],X_S[i_1],\dots,X_S[i_M]]^T$ is the known pilot with $\{i_m\}_{m=0}^{M-1}$ being the subcarrier indexes belonging to the pilots and $M$ being the total number of the pilots in one OFDM symbol, $\mathbf{x_d}=[X_S[j_0],X_S[j_1],\dots,X_S[j_Q]]^T$ is the desired data with $\{j_q\}_{q=0}^{Q-1}$ being the subcarrier indexes belonging to the data and $Q=N-M$ being the total number of data subcarriers in one OFDM symbol, and
$\mathbf{S_p}\in\mathbb{R}^{N\times M} $ and $ \mathbf{S_d}\in\mathbb{R}^{N\times Q} $ are the pilot-bearer matrix and data-bearer matrix, respectively, with their entries being
\begin{equation}
\begin{aligned}
\label{date-pilot-matrices}
&[\mathbf{S_d}]_{n, q}=
\begin{cases}
1, &  n=j_q\\
0, &  n\ne j_q 
\end{cases},\,
[\mathbf{S_p}]_{n,m}=\begin{cases}
1, &  n=i_m\\
0, &  n\ne i_m 
\end{cases},\\
&0\leq n\leq N-1, 0\leq m\leq M-1, 0\leq q\leq Q-1.
\end{aligned}
\end{equation}

The  data transmission stage is shown in Fig. \ref{Data_Transmission_Stage_1}. In this stage, $\mathbf{h_D}$ and $\mathbf{j}'_{\mathbf{I}}$ are estimated and the desired data $\mathbf{x_d}$ is detected. The designed approximation factor $K$ for the SI phase noise is set as $K=M-L_S$, which satisfies $L_S+K\leq N$ and promotes an unique solution for the estimations of $\mathbf{h_D}$ and $\mathbf{j}'_{\mathbf{I}}$\cite{4355336}. 
Then, we reconstruct one OFDM symbol transmitted from Node 2 as 
\begin{equation}
\label{reconstruction step}
\mathbf{\hat{x}_S}=\mathbf{S_px_p}+\mathbf{S_d\hat{x}_d}, 
\end{equation}
where $\mathbf{\hat{x}_d}$ is the detected desired data, and then the reconstructed symbol $\mathbf{\hat{x}_S}$ is fed back to the joint estimation part in the next iteration.
With the feedback step, the reconstructed symbol $\mathbf{\hat{x}_S}$ and the received signals on all subcarriers are exploited in the estimation step.
However, it should be ware that there might be data detection errors in the detected desired data $\mathbf{\hat{x}_d}$, which introduces certain noise in the reconstruction step.

\section{Performance Analysis}
\label{BER Bound in Data Transmission Stage}

In this section, the estimation and detection performance of the considered two-stage framework is analyzed. First, the MSE lower bound for the phase noise and channel estimation and the effective SINR upper bound for the received signal after SI cancellation are derived. Then, the BER lower bound for the detection of the desired data under the proposed framework is derived. 

Before the analysis, we present some assumptions adopted in this section: 
1) Since we focus on the joint phase noise estimation and data detection in the second data transmission stage, it is considered that perfect CSI for the SI channel $\mathbf{h_I}$ is obtained at the first stage; 
2) the estimation error and data detection error caused interference in the data transmission stage are simply treated as independent CSCG noises;
3) due to zero prior CSI at the transmitter, we do not consider the optimal power allocation among the carriers and thus each carrier of the OFDM symbol is allocated with identical power;
and 4) the data symbols, the pilots, the phase noise, the channel, and the CSCG noise are independent of each other.

\subsection{Lower Bound for Estimation MSE}
\label{Estimation Lower Bound with Data Detection Error}
In this subsection, we derive the lower bounds for the estimation MSE of $\mathbf{j_I'}$ and $\mathbf{h}_D$ under the considered framework shown in Fig. \ref{Data_Transmission_Stage_1} for the data transmission stage.

By utilizing the channel and phase noise approximation models given in \eqref{approximation model of SI phase noise 1} and \eqref{approximate model of SoI channel and SoI phase noise 1}, the received signal in \eqref{matrix form of receive signal 1} is rewritten as
\begin{equation}
\label{matrix form of received signal 2}
\begin{split}
\mathbf{y}&=\mathbf{T_I}\mathbf{j_I}+\mathbf{J_SH_SX_S}+\mathbf{w}
\\&=\mathbf{T_I}\mathbf{S_I}\mathbf{j_I'}
+\mathbf{H_Dx_S}+\mathbf{e},
\end{split}
\end{equation}
where the $m$-th row and $n$-th column entry of $\mathbf{T_I}$ is given as $[\mathbf{T_I}]_{m, n}=H_I[m-n]X_I[m-n],\, 0\leq m\leq N-1, 0\leq n \leq N-1,\dots,N-1$, and 
\begin{equation}
\label{n 1}
\mathbf{e}=\mathbf{T_I}(\mathbf{j_I}-\mathbf{S_I}\mathbf{j_I'})+\mathbf{(J_SH_S-H_D)x_S}+\mathbf{w}
\end{equation} 
is the combination of the CSCG noise $\mathbf{w}$ and the modeling errors introduced in \eqref{approximation model of SI phase noise 1} and \eqref{approximate model of SoI channel and SoI phase noise 1}. The power of $\mathbf{e}$ is derived as
\begin{equation}
\label{sigma n 1}
\begin{split}
\sigma_e^2=\mathbb{E}[\Vert{\mathbf{e}\Vert}^2]=\lambda_IE_{h_I}E_I+\lambda_SE_{h_S}E_S+\sigma_{w}^2,
\end{split}
\end{equation}
where $E_S=\mathrm{E}[\Vert\mathbf{x_S}\Vert^2]$, $E_I=\mathrm{E}[\Vert\mathbf{x_I}\Vert^2]$, $E_{h_I}=\mathrm{E}[\Vert\mathbf{h_I}\Vert^2]$,  and $E_{h_S}=\mathrm{E}[\Vert\mathbf{h_S}\Vert^2]$ are the power values of the transmitted symbol, the SI symbol, the SI channel response, and the SoI channel response, respectively, and $\lambda_I=\sum_{k=K}^{N-1}\mathbb{E}[|J_I[k]|^2]$ and $\lambda_S=\sum_{k=1}^{N-1}\mathbb{E}[|J_S[k]|^2]$ are the power values of the residual SI phase noise and the residual SoI phase noise, respectively. 
It is observed from \eqref{sigma n 1} that the existence of phase noise has introduced interference, i.e., the model approximation errors $\lambda_IE_{h_I}E_I$ and $\lambda_SE_{h_S}E_S$ ($\lambda_I$ and $\lambda_S$ are calculated in Appendix \ref{Power of the residual phase noise} in details), which degrade the estimation performance.
Besides, the power values $\lambda_I$ and $\lambda_S$ are determined by the model of phase noise, which we will discuss in the simulation and Appendix \ref{Power of the residual phase noise}.

\begin{Proposition}
\label{lower bound of the parameter estimation errors 1}	
If the desired data vector $\mathbf{x_d}$ is detected as $\mathbf{\hat{x}_{d_{n-1}}}$ in the $(n-1)$-th iteration of the data transmission stage, $n>1$, then the estimation MSE of $\mathbf{j_I'}$ and $\mathbf{h_D}$ in the $n$-th iteration are lower bounded as
	\begin{equation}
	\label{lower bound of the parameter lower bound of the parameter estimation errors 1}
	\begin{aligned}
	C_n\ge&\frac{(M-L_S)(\sigma_e^2+(1-\lambda_S)E_{h_S}d(n-1))}{NE_{hI}E_I},\\
	D_n\ge&\frac{L_S(\sigma_e^2+(1-\lambda_S)E_{h_S}d(n-1))}{NE_S},
	\end{aligned}
	\end{equation} 
	where $C_{n}=\mathbb{E}\left[\Vert{\mathbf{j'_I}-\mathbf{\hat{j}}'_{\mathbf{I}_{n}}\Vert}^2\right]$ is the estimation MSE of $\mathbf{j_I'}$, with $\mathbf{\hat{j}}'_{\mathbf{I}_{n}}$ being the estimation of $\mathbf{j_I'}$ in the $n$-th iteration, $D_{n}=\mathbb{E}[\Vert{\mathbf{h_D}-\mathbf{\hat{h}}_{\mathbf{D}_{n}}\Vert}^2]$ is the estimation MSE for $\mathbf{h_D}$, with $\mathbf{\hat{h}}_{\mathbf{D}_{n}}$ being the estimation of $\mathbf{h_D}$ in the $n$-th iteration, and $d(n-1)=\Vert{\mathbf{x_d}-\mathbf{\hat{x}_{d_{n-1}}}\Vert}^2$ is the square error between the desired data vector $\mathbf{x_d}$ and its data detection result $\mathbf{\hat{x}_{d_{n-1}}}$
	in the $(n-1)$-th iteration.
	\begin{IEEEproof}
		See Appendix \ref{CRLB Of Parameter Estimation}.	
	\end{IEEEproof}
\end{Proposition}

Notice that the data detection result $\mathbf{\hat{x}}_{\mathbf{d}_{n-1}}$ of the $(n-1)$-th iteration is utilized in the estimation process of the $n$-th iteration.  Thus, the lower bounds for the estimation MSE of $\mathbf{j_I'}$ and $\mathbf{h_D}$ are related to both the noise $\mathbf{e}$ which is independent across different iterations and the data detection square error $d(n-1)$ which is varying over iterations.
	
\subsection{Upper Bound for Effective SINR}
\label{Effective SINR in Each Iteration}
In this section, we derive the upper bound for the effective SINR of the received signal after SI cancellation in each iteration. 

According to the considered framework shown in Fig. \ref{Data_Transmission_Stage_1}, in the $n$-th iteration of the data transmission stage, $n>1$, we first utilize the estimated SI channel from the SI channel estimation stage and the estimated SI phase noise $\mathbf{\hat{j}}'_{\mathbf{I}_{n}}$, to reconstruct the SI signal. Then, the reconstructed SI signal is used for the SI cancellation to recover the SoI from the received signal $\mathbf{y}$. Using \eqref{matrix form of received signal 2} and the SI phase noise model approximation shown in section \ref{SI Channel Estimation Stage}, the received signal after SI cancellation is given as 
\begin{equation}
	\label{residual data-located signal 2}
	\begin{split}
		\mathbf{r}_{n}&= \mathbf{S}_{\mathbf{d}}^T(\mathbf{y}-\mathbf{T_I}\mathbf{S_I}\mathbf{\hat{j}}'_{\mathbf{I}_{n}})\\
		&= \mathbf{s}+\mathbf{S}_{\mathbf{d}}^T\mathbf{T_I}\mathbf{S_I}(\mathbf{j'_I}-\mathbf{\hat{j}}'_{\mathbf{I}_n})+\mathbf{S}_{\mathbf{d}}^T\mathbf{e},
	\end{split}
\end{equation}
where $\mathbf{r}_{n}$  is the received signal after SI cancellation in the $n$-th iteration, $\mathbf{T_I}\mathbf{S_I}\mathbf{\hat{j}}'_{\mathbf{I}_{n}}$ is the reconstructed SI signal, $\mathbf{s}=\mathrm{diag}\{\mathbf{S}_{\mathbf{d}}^T\mathbf{F_Sh_D}\}\mathbf{x_d}$ is the SoI related to the desired data $\mathbf{x_d}$, and $\mathbf{r}_{n}-\mathbf{s}$ denotes the residual SI. Based on the estimated SoI channel $\mathbf{\hat{h}}_{\mathbf{D}_{n}}$, the desired data $\mathbf{x_d}$ is detected from $\mathbf{r}_{n}$.

The detection performance of the desired data $\mathbf{x_d}$ is determined by the effective SNR of $\mathbf{r}_n$. By using the MSE lower bounds derived in \eqref{lower bound of the parameter lower bound of the parameter estimation errors 1} for $\mathbf{j_I'}$ and $\mathbf{h_D}$, the upper bound for the average effective SINR of $\mathbf{r}_n$ is derived in the following proposition.
\begin{Proposition}
\label{Effective SINR 1}
Define $\overline{\gamma}(n)$ as the average effective SINR of the receive signal $\mathbf{r}_n$ after SI cancellation in the $n$-th iteration, then $\overline{\gamma}(n)$ is upper bounded as
\begin{equation}
	\label{effective SINR 4}
	\begin{split}
		\overline{\gamma}(n)\leq \frac{(1-\lambda_S)E_{h_S}E_S}{(1+\frac{M}{N})\sigma_e^2+\frac{MQ}{N^2}(1-\lambda_S)E_{h_S}\overline{d}(n-1)},
	\end{split}	
\end{equation}
where $E_S$, $E_I$, $E_{h_I}$, $E_{h_S}$, $\lambda_I$ and $\lambda_S$ are defined in \eqref{sigma n 1}, and $\overline{d}(n-1)=|X_S-\hat{X}_S^{n-1}|^2$ is the square data detection error in one OFDM subcarrier,
with $X_S$ being the desired data symbol and $\hat{X}_S^{n-1}$ being the detected $X_S$ in the $(n-1)$-th iteration.	
\begin{IEEEproof}
See Appendix \ref{Proof of the closed-form of SINR}.
\end{IEEEproof}
\end{Proposition}
	
Notice that $\overline{d}(n-1)$ is related to the adopted modulation scheme of the desired data. First, we start with a simple case that the binary-phase-shift-keying (BPSK) modulation is adopted. Thus, we have $X_S\in \{ \sqrt{E_S/N}, -\sqrt{E_S/N}\}$, and $\overline{d}(n-1)$ has two possible values corresponding to the two possible data detection results, i.e.,
\begin{equation}
	\label{data detection errors 2}
	\begin{split}
		\begin{cases}
			S^0_{n-1}: \overline{d}(n-1)=0\\
			S^1_{n-1}: \overline{d}(n-1)=4\frac{E_S}{N},
		\end{cases}
	\end{split}
\end{equation}
where $S^0_{n-1}$ denotes the correct detection result in the $(n-1)$-th iteration, i.e., $\hat{X}_S^{n-1}=X_S$, and $S^1_{n-1}$ denotes the false detection result in the $(n-1)$-th iteration, i.e., $\hat{X}_S^{n-1}=-X_S$. By substituting \eqref{data detection errors 2} into \eqref{effective SINR 4}, the upper bound for the effective SINR in the $n$-th iteration is derived as 
\begin{equation}
	\label{effective SINR 5}
	\begin{split}
		\begin{cases}
			S^0_{n-1}: \overline{\gamma}(n)\leq \gamma_0=\dfrac{(1-\lambda_S)E_SE_{h_S}}{(1+\frac{M}{N})\sigma_e^2}\\
			S^1_{n-1}: \overline{\gamma}(n)\leq \gamma_1,
		\end{cases}
	\end{split}
\end{equation}
where $\gamma_0$ denotes the effective SINR upper bound for case $S_{n-1}^0$, and                                                
\begin{equation}
	\label{gamma1}
	\gamma_1=\dfrac{(1-\lambda_S)E_{h_S}E_S}{(1+\frac{M}{N})\sigma_e^2+4\frac{MQ}{N^2}(1-\lambda_S)E_{h_S}E_S},
\end{equation}
denotes the effective SINR upper bound for case $S_{n-1}^1$.

For general modulation schemes, we consider the case that the desired data symbol $X_S\in \{a_1, a_2, \cdots, a_T\}$, with $p(X_S=a_i)=1/T$, $i=1, 2, \cdots, T$. 
Similar to the above analysis for BPSK modulation, the square detection error $\overline{d}(n-1)$ has $T^2$ possible values corresponding to $T^2$ possible data detection results. 
For the case that $\hat{X}_S^{n-1}=a_j$ with the desired data $X_S=a_i$, the corresponding square detection error is given as
	\begin{equation}
	\label{data detection errors 3}
	\overline{d}(n-1)=|a_i-a_j|^2,\,
	i,j= 1,2,\cdots, T.
	\end{equation}
	Then, by substituting \eqref{data detection errors 3} into \eqref{effective SINR 4}, we derive the upper bound for the effective SINR of the $n$-th iteration with different detection results from the $(n-1)$ iteration. For the case that $\hat{X}_S^{n-1}=a_j$ with $X_S=a_i$, the effective SINR of the n-th iteration is upper bounded as
	\begin{equation}
	\label{effective SINR 7}
	\overline{\gamma}(n)\leq \gamma_{i,j},\,
	i,j= 1,2,\cdots, T,
	\end{equation}
	where	
	$$ \gamma_{i,j}=\frac{(1-\lambda_S)E_{h_S}E_S}{(1+\frac{M}{N})\sigma_e^2+\frac{MQ}{N^2}(1-\lambda_S)E_{h_S}|a_i-a_j|^2}$$
    is the corresponding effective SINR upper bound for the case $\hat{X}_S^{n-1}=a_j$ with $X_S=a_i$.
	
\subsection{Lower Bound for Data Transmission BER}
\label{BER Lower Bound in the Data Transmission Stage}	

In this section, by utilizing the effective SINR bound derived in section \ref{Effective SINR in Each Iteration}, we derive the lower bound for the BER in the data transmission stage. Similar to the analysis of the SINR upper bound, the BER lower bound is also related to the adopted modulation scheme of the desired data.

We first take the BPSK modulation under the Rayleigh fading channels as an example to analyze the corresponding BER lower bound. For the Rayleigh fading channel, the BER of the data transmission with BPSK modulation can be expressed as a function of the SINR\cite{1053806}, i.e. 
\begin{equation}
	\label{Rayleigh fading channel BER 1}
	P=f(x)=\frac{1}{2}\left(1-\sqrt{\frac{x}{1+x}}\right),
\end{equation} 
where $P$ is the BER and $x$ denotes the value of the effective SINR, with $x\ge 0$.
\begin{Proposition}
\label{BER Bound 1}
For the consider two-stage framework under the Rayleigh fading channels, define $P_b$ as the BER of the desired data with BPSK modulation, and $P_b$ is lower bounded as
\begin{equation}
	\label{BER Bound 2}
	\begin{split}
		P_{b}\ge \frac{f(\gamma_0)}{1+f(\gamma_0)-f(\gamma_1)},
	\end{split}
\end{equation}
where $\gamma_0$ and $\gamma_1$ are defined in \eqref{effective SINR 5} as the effective SINR upper bounds for BPSK modulation.
\begin{IEEEproof}
See Appendix \ref{Appendix: BER Bound}.	
\end{IEEEproof}
\end{Proposition}

Then, we investigate the universal BER lower bound valid for arbitrary modulation schemes. Consider that the desired data symbol $X_S\in \{a_1, a_2, \cdots, a_T\}$, with $p(X_S=a_i)=1/T$, $i=1, 2, \cdots, T$. 
Similar to \eqref{Rayleigh fading channel BER 1}, the probability of $T^2$ possible detection cases can be denoted as $T^2$ functions of the corresponding effective SINR. That is, for the data detection result $\hat{X}_S=a_j$ with the desired data $X_S=a_i$, the corresponding detection probability can be expressed as a deterministic function of the effective SINR, i.e. 
\begin{equation}
\label{conditional probability 2}
	    p(\hat{X}_S=a_j|X_S=a_i)= f_{i,j}(x),
\end{equation}
where $x$ is the value of the corresponding effective SINR for the case $\hat{X}_S=a_j$ with $X_S=a_i$, and the expression of the function $f_ij(\cdot)$ is determined by the adopted modulation scheme and the data transmission channel.
Here, we consider the case that $f_{i,j}(x)$ is monotonically decreasing with respect to $x$ for $i\neq j$, and for $i=j$, $f_{i,i}(x)$ is monotonically increasing. This assumption is valid due to the intuition that when the effective SINR value denoted by x becomes larger, the correct detection probability  $f_{i, i}(x)$ should be increasing, and the false detection probability $f_{i,j}(x)\ (i\neq j)$ should be decreasing.

In the $n$-th iteration, for the data detection result $\hat{X}_S=a_j$ with the desired data $X_S=a_i$, the corresponding detection probability can be computed as
\begin{equation}
\label{Symbol error rate 1}
p(\hat{X}_S^{n}=a_j|X_S=a_i)=
\sum_{q=1}^{T} p(\hat{X}_S^{n}=a_j|X_S=a_i, \hat{X}_S^{n-1}=a_q)p(\hat{X}_S^{n-1}=a_q|X_S=a_i).
\end{equation}	
Define  $\overline{\gamma}(n)|_{X_S=a_i, \hat{X}_S^{n-1}=a_q}$ as the effective SINR of $n$-th iteration when the case $\hat{X}_S^{n-1}=a_q$ with $X_S=a_i$ occurs and substitute it into  \eqref{conditional probability 2}, it follows
	\begin{equation}
	\label{effective SINR 8}
	\begin{split}
	p(\hat{X}_S^{n}=a_j|X_S=a_i, \hat{X}_S^{n-1}=a_q)=f_{i,j}(\overline{\gamma}(n)|_{X_S=a_i, \hat{X}_S^{n-1}=a_q}).
	\end{split}
	\end{equation}
Then, by substituting the derived effective SINR upper bounds for $T^2$ data detection results in \eqref{effective SINR 7} into \eqref{effective SINR 8}, it follows
	\begin{equation}
	\label{effective SINR 9}
	\begin{split}
	\begin{cases}
	p(\hat{X}_S^{n}=a_j|X_S=a_i, \hat{X}_S^{n-1}=a_q)\ge f_{i,j}(\gamma_{i,q}), i\neq j,\\
	p(\hat{X}_S^{n}=a_j|X_S=a_i, \hat{X}_S^{n-1}=a_q)\leq f_{i,j}(\gamma_{i,q}), i=j.
	\end{cases}
	\end{split}
	\end{equation}
	
Define $P_{i,j}=\lim\limits_{n \to \infty}p(\hat{X}_S^{n}=a_j|X_S=a_i)$ as the asymptotic detection probability of the desired data in data transmission stage. Similar to the analysis for the BER of BPSK modulation in \eqref{BER 1}, we substitute \eqref{effective SINR 9} into \eqref{Symbol error rate 1}, and $\{P_{i,j}\}_{i,j=1}^T$ should satisfy the following constrains, i,e.,
\begin{equation}
\label{Symbol error rate 2}
\begin{split}
\begin{cases}
P_{i,j}\ge \sum_{q=1}^{T}f_{i,j}(\gamma_{i,q})P_{i,q},\,
i\neq j\\
P_{i,j}\leq \sum_{q=1}^{T}f_{i,j}(\gamma_{i,q})P_{i,q},\,
i= j\\
\sum_{i=1}^{T}\sum_{j=1}^{T} P_{i,j}=1.
\end{cases}
\end{split}
\end{equation}
Define $P_b$ as the BER of the desired data in data transmission stage, and then $P_b$  is determined by $\{P_{i,j}\}_{i,j=1}^T$, i.e.,
	\begin{equation}
	\label{BER 3}
	\begin{split}
	P_b=\frac{1}{\log_2T}\sum_{i=1}^{T}\sum_{j=1,j\neq i}^{T} \alpha(i,j)P_{i,j},
	\end{split}
	\end{equation} 
where $\alpha(i, j)$ is the number of bit errors caused by the data detection result $\hat{X}_S=a_j$ with the desired data $X_S=a_i$. Combing \eqref{Symbol error rate 2} and \eqref{BER 3}, we conclude that $P_b$ is lower bounded, and the lower bound is determined by $\{P_{i,j}\}_{i,j=1}^T$ satisfying the constrains in \eqref{Symbol error rate 2}.
		
\section{Numerical and Simulation Results} 
\label{Simulation Analysis}
This section presents the simulations for the considered two-stage joint SI cancellation and data detection framework and its theoretical BER bound. 
For the considered two-stage framework, we adopt the least square estimation and the maximum likelihood detection algorithms to verify its performance \cite{9013150}.

Two transceivers are considered to simultaneously transmit and receive the BPSK data in the same frequency-band. We set the number of OFDM subcarriers $N$ as 1024, the length of cyclic prefix as 32,
the subcarrier spacing $f_{sub}$ as 10 kHz, and the sample time $T_s=1/(f_{sub}\cdot N)=10^{-7}\,$s. The number of multipath SoI channel taps is $L_S=20$, and the number of pilots in one OFDM symbol is $M=40$.
The phase noise at both Nodes 1 and 2 are modeled as independent identically distributed Wiener processes \cite{6937196}, and their variance is $\sigma_{\theta}^2=4\pi \Delta f$, where the relative bandwidth $\Delta f$ of the phase noise reflects the quality of the oscillator \cite{6937196}.

\subsection{BER versus SNR and Iterations}
\begin{figure}[htbp]
	\setlength{\abovecaptionskip}{6pt}
	\centering
	\begin{minipage}[t]{230pt}
		\centering
		\includegraphics[width=230pt]{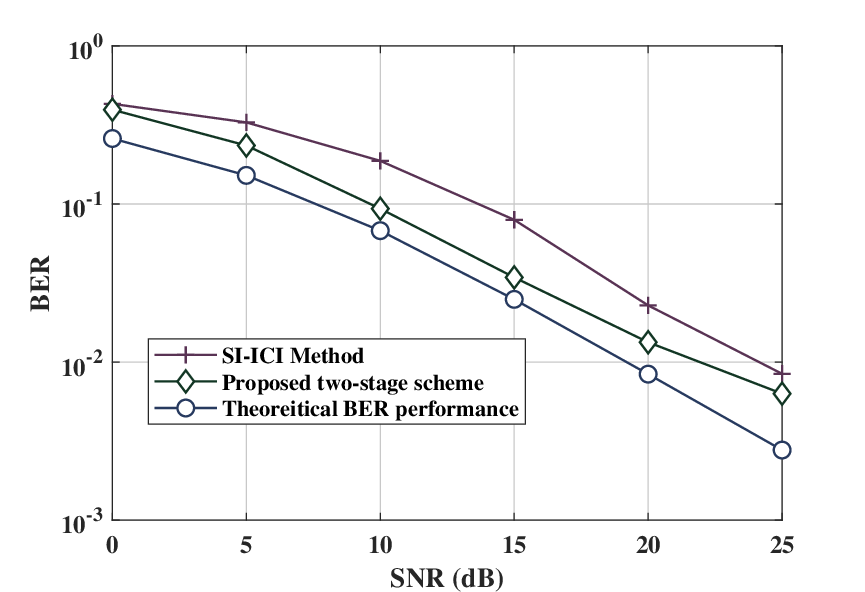}
		\caption{BER versus SNR under different SI cancellation methods with INR = 40\ dB and $\Delta f=10^{-4}$.}
		\label{state of the art}
	\end{minipage}
\end{figure}

Fig. \ref{state of the art} compares the BER performance of the considered two-stage framework and an SI cancellation  method with ICI suppression (SC-ICI) in \cite{6937196}. The SC-ICI method suppresses both the CPE and ICI of the phase noise and treats the SoI as the interference in the SI cancellation process. Thus, the SI cancellation ability of SC-ICI is limited by the power of SoI. 
From this figure, it is first observed that the BER performance of the considered two-stage framework is close to and consistent with its theoretical BER lower bound derived in \eqref{BER Bound 2}, which verifies our analysis in section IV and validates the feasibility of this framework.
Then, it is further observed that the considered two-stage scheme outperforms the SC-ICI method under different SNR scenarios, due to the fact that the two-stage framework considers the joint SI cancellation and SoI data detection, which suppresses the SoI interference in the SI cancellation process, and thus our method is better than the SC-ICI method under certain scenarios.

\begin{figure}[htbp]
	\setlength{\abovecaptionskip}{6pt}
	\centering
	\begin{minipage}[t]{230pt}
		\centering
		\includegraphics[width=230pt]{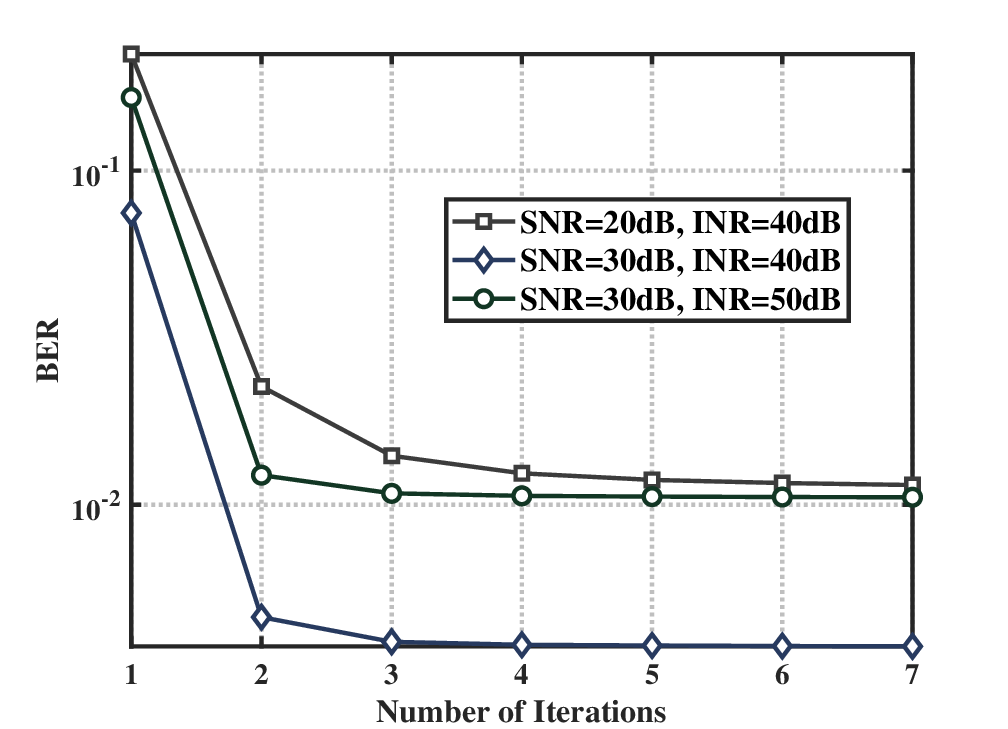}
		\caption{BER versus numbers of iterations under different SNR and INR with $\Delta f=10^{-4}$.}
		\label{BER_1}
	\end{minipage}
	\hfill%
	\begin{minipage}[t]{230pt}
		\centering
		\includegraphics[width=230pt]{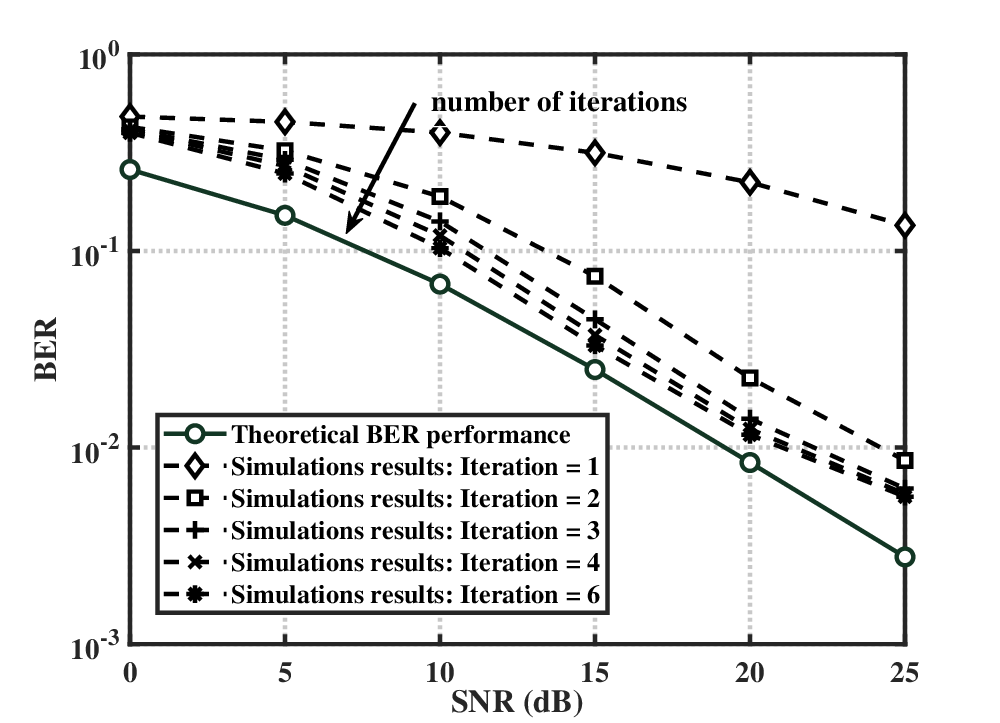}
		\caption{BER versus SNR under different numbers of iterations with INR = 40\ dB and $\Delta f=10^{-4}$.}
		\label{BER_2}
	\end{minipage}
\end{figure}
The BER performance of the considered two-stage scheme versus numbers of iterations is shown in Fig. \ref{BER_1}. From this figure, it is observed that in all scenarios with different SNRs and INRs, the BER of the considered two-stage scheme decreases and quickly converges after 4 to 6 iterations. Besides, we notice that the converged BER performance for the scenarios of $\mathrm{SNR}=20 \ \mathrm{dB}$/$\mathrm{INR}=40\ \mathrm{dB}$ and $\mathrm{SNR}=30\ \mathrm{dB}$/$\mathrm{INR}=50\ \mathrm{dB}$ are almost identical, which implies that the BER performance is related to the ratio between the SNR and INR.
We also noticed that the converged BER for the scenario of $\mathrm{SNR}=30\ \mathrm{dB}$/$\mathrm{INR}=40\ \mathrm{dB}$ is smaller than the other two scenarios, which implies that lager ratio between the SNR and INR leads to smaller BER.

Fig. \ref{BER_2} plots the BER performance of the considered two-stage scheme as a function of SNR under different numbers of iterations. 
From this figure, it is observed that as the SNR increases from 0 dB to 25 dB, the BER performance with only one iteration slowly drops from $10^{-0.5}$ to $10^{-1}$, and the BER performance with 6 iterations, which is close to and consistent with the theoretical BER lowered bound derived in \eqref{BER Bound 2}, significantly drops from $10^{-0.5}$ to $10^{-2.2}$. 
It is also observed from this figure that only 6 iterations are sufficient to achieve almost all the performance gain.

\subsection{Estimation MSE}
\begin{figure}[htbp]
	\setlength{\abovecaptionskip}{6pt}
	\centering
	\begin{minipage}[t]{230pt}
		\centering
		\includegraphics[width=230pt]{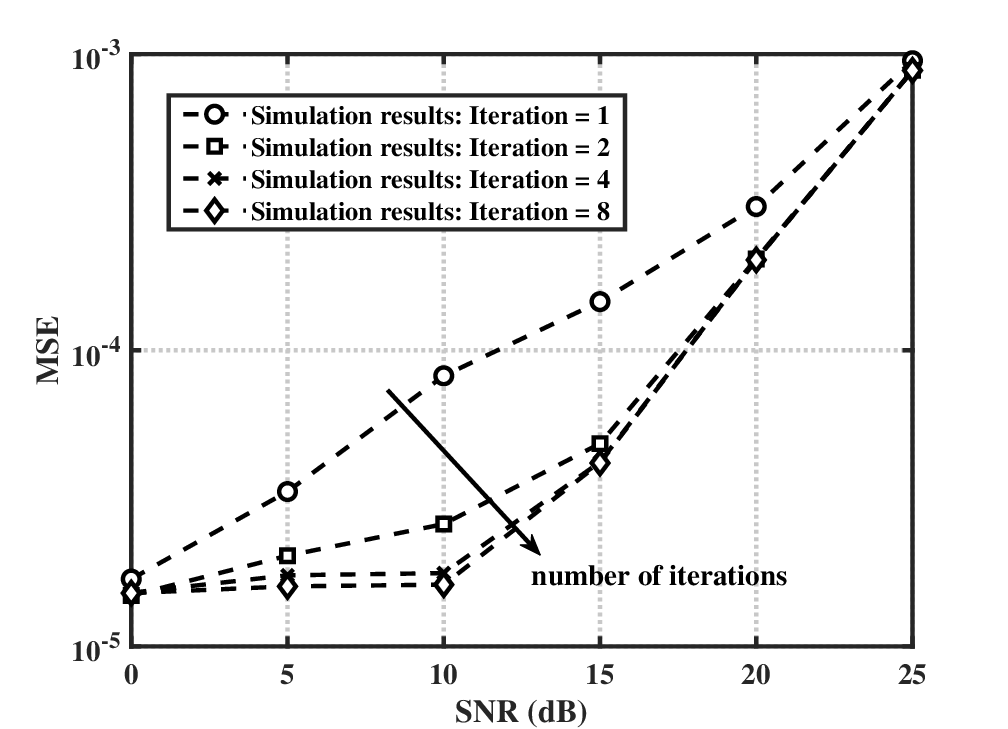}
		\caption{Estimation MSE of the SI phase noise versus SNR under different numbers of iterations, with $\mathrm{INR}=40\ \mathrm{dB}$ and $\Delta f=10^{-4}$.}
		\label{data1_MSE_1}
	\end{minipage}
	\hfill%
	\begin{minipage}[t]{230pt}
		\centering
		\includegraphics[width=230pt]{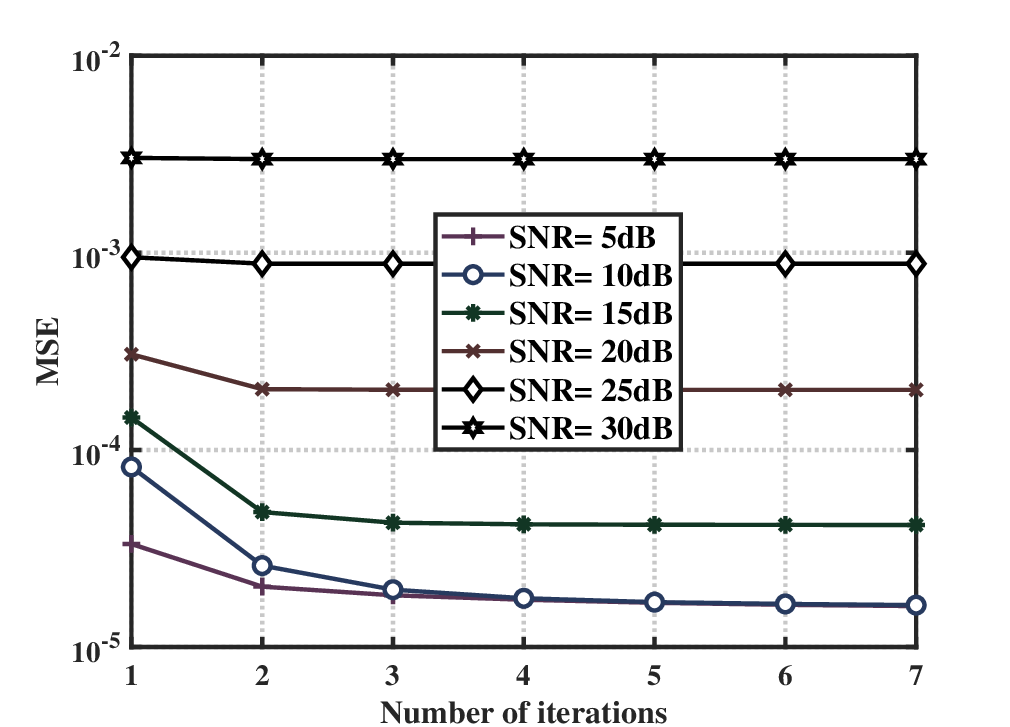}
		\caption{Estimation MSE of the SI phase noise versus numbers of iterations under different SNR, with $\mathrm{INR}=40\ \mathrm{dB}$ and $\Delta f=10^{-4}$.}
		\label{data1_MSE_2}
	\end{minipage}
\end{figure}
Fig. \ref{data1_MSE_1} plots the estimation MSE of the SI phase noise as a function of SNR under different numbers of iterations. From the plot, it is observed that as the SNR increases from 0 dB to 25 dB, the MSE monotonically increases from about $10^{-4.8}$ to $10^{-3}$. This phenomena can be explained by the analysis in section \ref{Effective SINR in Each Iteration} and Appendix \ref{CRLB Of Parameter Estimation}: According to \eqref{power of z 1} and \eqref{C 1}, the SI phase noise estimation MSE lower bound is proportionally related to the power of the noise $\mathbf{e}$ defined in \eqref{sigma n 1}, i.e., $\sigma_e^2=\lambda_IE_{h_I}E_I+\lambda_SE_{h_S}E_S+\sigma_{w}^2$, and thus when the SNR becomes larger, $\lambda_SE_{h_S}E_S$ becomes larger and the estimation MSE of the SI phase noise tends to increase.

The estimation MSE of the SI phase noise is shown in Fig. \ref{data1_MSE_2}, with different numbers of iterations and SNRs. From this figure, it is observed that in all scenarios with different SNRs, the MSE tends to decrease and converge with the increasing of iterations. 
Besides, we notice that for the case with larger SNR, the SI phase noise estimation performance, in terms of the estimation MSE, has smaller gain from the iterations. For example, under the scenario with SNR $=30$ dB, the converged MSE is very close to the MSE with only one iteration. In other words, the estimation performance cannot benefit from the iterations under the scenario with high SNR.

\subsection{Phases Noise Effect}
\begin{figure}[htbp]
	\setlength{\abovecaptionskip}{6pt}
	\centering
	\begin{minipage}[t]{230pt}
		\centering
		\includegraphics[width=230pt]{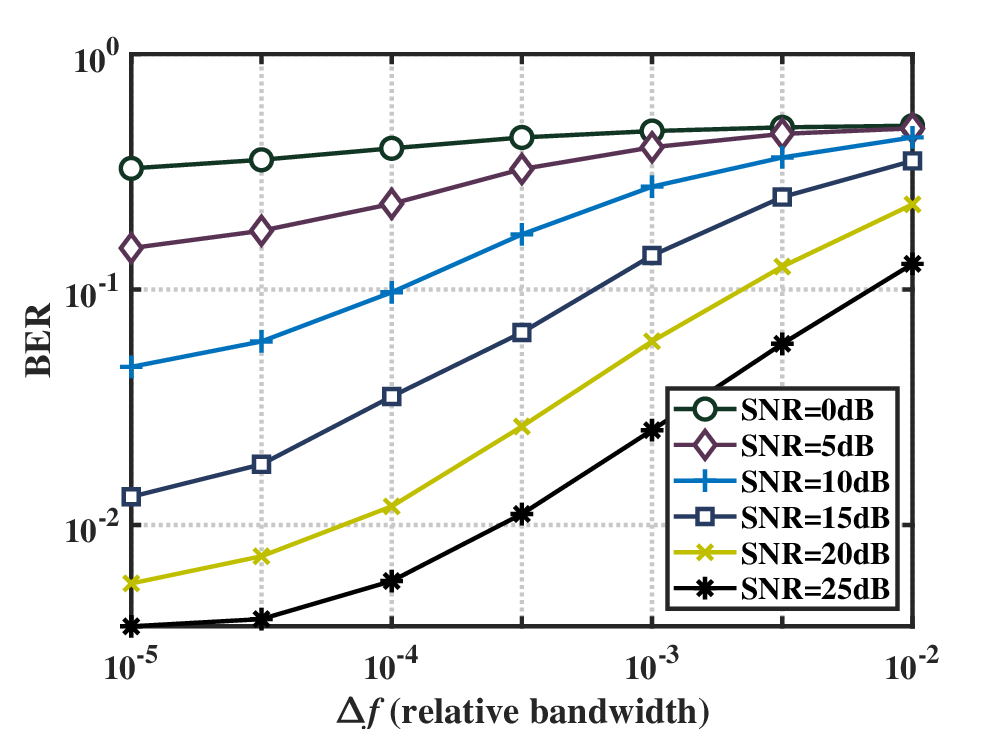}
		\caption{BER versus $\Delta f$ under different SNRs with $\mathrm{INR}=40$ dB.}
		\label{BER_SNR_Delta_1}
	\end{minipage}
	\hfill%
	\begin{minipage}[t]{230pt}
		\centering
		\includegraphics[width=230pt]{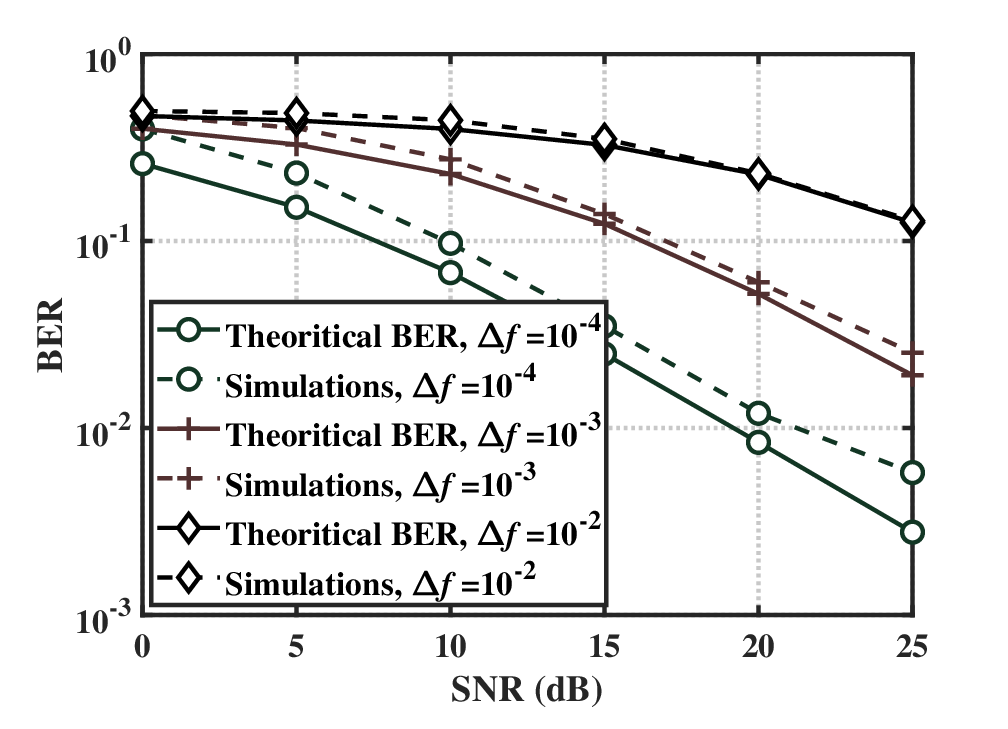}
		\caption{BER versus SNR under different $\Delta f$ with INR = $40$ dB.}
		\label{BER_Delta_SNR_1}
	\end{minipage}
\end{figure}

\begin{figure}[htbp]
	\setlength{\abovecaptionskip}{6pt}
	\centering
	\begin{minipage}[t]{230pt}
		\centering
		\includegraphics[width=230pt]{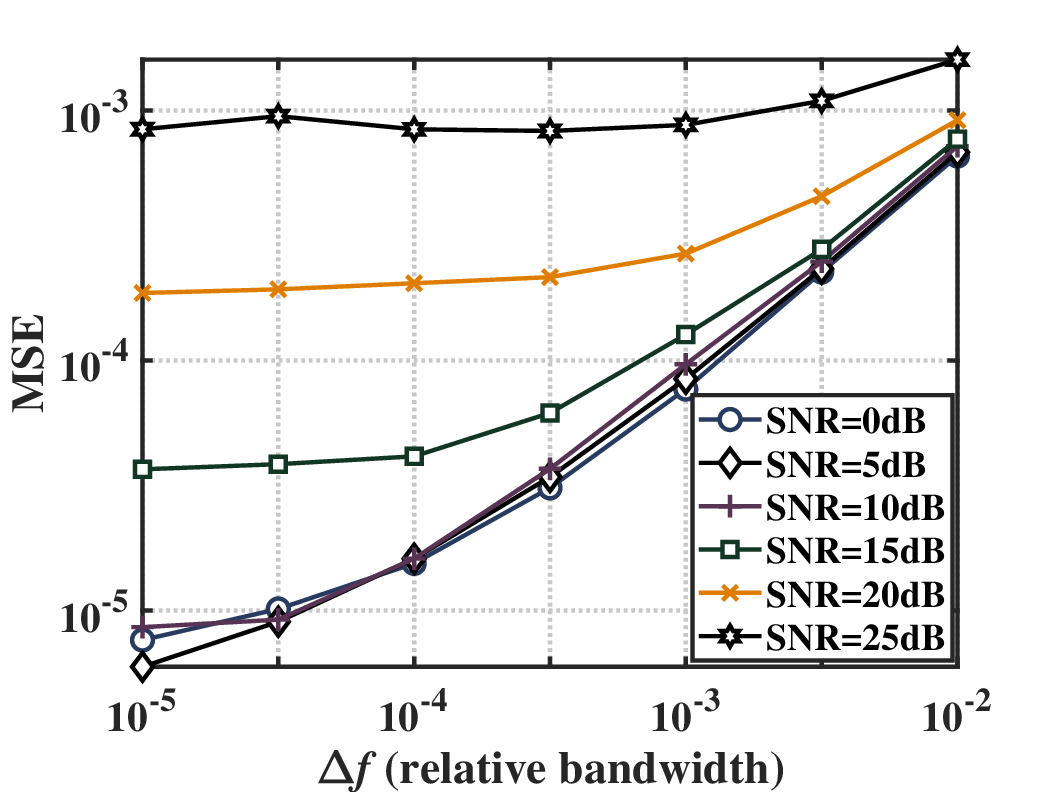}
		\caption{Estimation MSE of the SI phase noise versus $\Delta f$ under different SNR with INR = $40$ dB.}
		\label{MSE_Delta_SNR_1}
	\end{minipage}
\end{figure}
Fig. \ref{BER_SNR_Delta_1} shows the simulation results for the BER of the considered scheme versus the relative bandwidth $\Delta f$ under different SNRs. In all scenarios with different SNRs, as $\Delta f$ increases from $10^{-5}$ to $10^{-2}$, the BER significantly increases. 
This phenomena can be interpreted by the analysis of the SINR upper bound in section \ref{Effective SINR in Each Iteration}: The effective SINR is limited by the power of the residual SI phase noise, which increases monotonically with $\Delta f$ according to the analysis in appendix \ref{Power of the residual phase noise}. Therefore, when $\Delta f$ increases, the power of the residual SI phase noise increases with it, the effective SINR decreases with the increase of the power of the residual SI phase nose, and the BER increases with the decrease of the SINR.
It is concluded that $\Delta f$ is a key parameter which influences the data detection performance of the considered scheme.

Fig. \ref{BER_Delta_SNR_1} compares the BER of the considered scheme with its theoretical BER lower bound derived in \eqref{BER Bound 2}, under different SNRs and $\Delta f$. From this figure, it is observed that in all scenarios with different $\Delta f$, the BER performance of the considered scheme are all highly close to and consistent with the theoretical BER lower bound. Besides, we can also observe that when $\Delta f$ is smaller, the gap between the simulations and the corresponding lower bound is also smaller.

Fig. \ref{MSE_Delta_SNR_1} shows the estimation MSE of the SI phase noise versus $\Delta f$ under different SNRs. It is observed from this figure that in all scenarios with different SNRs, as $\Delta f$ increases from $10^{-5}$ to $10^{-2}$, the estimation MSE significantly increases. Besides, we can also observe that in the scenario with larger SNR, the estimation MSE is less sensitive to the variation of $\Delta f$. For example, when $\Delta f$ increases from $10^{-5}$ to $10^{-2}$, for the case with SNR = $0$ dB, the MSE increases from about $10^{-5}$ to $10^{-3}$, while for the case with SNR = $25$ dB, the MSE only increases from about $10^{-3.1}$ to about $10^{-2.8}$.

\section{Conclusion}
\label{Conclusion}
In this paper, a two-stage framework is considered to jointly suppress the strong SI and detect the desired data for a FD two-way OFDM system.
The first stage is the SI channel estimation, wherein the joint channel and phase noise estimation is proposed to obtain the estimation of SI channel. The second stage is the data transmission, wherein an iterative structure with feed back is proposed to jointly detect the desired data and estimate the unknown SoI channel and SI phase noise. The estimation MSE lower bounds for the channel and phase noise in each iteration are first derived.  Then, by subsisting the MSE lower bounds into the SINR expression that related to the channel and phase noise estimation MSE, the upper bound of the SINR in each iteration is computed.
Finally, by exploiting the transition information of the detection errors between two adjacent iterations, we derive the closed-form expression of the BER lower bound for BPSK modulation and then extend the analysis to a more general case with arbitrary modulation schemes. The derived universal BER lower bound is also independent of any specific estimation and detection methods. In the future, we will further investigate the joint SI cancellation and data detection for the multi-antenna FD systems.

\appendices
\section{Proof of Proposition \ref{lower bound of the parameter estimation errors 1}}
\label{CRLB Of Parameter Estimation}
According to the considered framework shown in Fig. \ref{Data_Transmission_Stage_1}, in the $n$-th iteration of the data transmission stage, $n>1$, we utilize the known $\mathbf{x_I}$, $\mathbf{h_I}$, and the reconstructed symbol $\mathbf{\hat{x}_{\mathbf{S}_{n-1}}}=\mathbf{S_px_{p}}+\mathbf{S_d}\mathbf{\hat{x}}_{\mathbf{d}_{n-1}}$, where $\mathbf{\hat{x}}_{\mathbf{d}_{n-1}}$ is the detection results of $\mathbf{x_d}$ in the $(n-1)$-th iteration, to estimate $\mathbf{j_I'}$ and $\mathbf{h_D}$. Thus, it indicates that the estimation performance in the $n$-th iteration is affected by the data detection result in the $(n-1)$-th iteration.

In the $n$-th iteration, by using \eqref{x_S,X_p,X_d} and \eqref{reconstruction step},
the received signal in \eqref{matrix form of received signal 2} is rewritten as
\begin{equation}
\begin{split}
\label{received signal with data detection errors 1}
\mathbf{y}=&\mathbf{T_I}\mathbf{S_I}\mathbf{j_I'}
+\mathbf{H_D\mathbf{\hat{x}}_{\mathbf{S}_{n-1}}}+\mathbf{H_D}(\mathbf{x_S}-\mathbf{\hat{x}}_{\mathbf{S}_{n-1}})+\mathbf{e}\\
=&\mathbf{T_I}\mathbf{S_I}\mathbf{j_I'}
+\mathbf{H_D\mathbf{\hat{x}}_{\mathbf{S}_{n-1}}}+\mathbf{H_DS_d}(\mathbf{x_d}-\mathbf{\hat{x}}_{\mathbf{d}_{n-1}})+\mathbf{e}.
\end{split}
\end{equation}
It is easy to observe that the detection error between the desired data symbol $\mathbf{x_d}$ and the corresponding detection result $\mathbf{\hat{x}}_{\mathbf{d}_{n-1}}$ from $(n-1)$-th iteration will introduce certain noise in the estimation step of the n-th iteration.
Thus, we rewrite the received signal in \eqref{received signal with data detection errors 1} as 
\begin{equation}
\label{received signal with data detection errors 2}
\mathbf{y}=\mathbf{T_I}\mathbf{S_I}\mathbf{j_I'}
+\mathbf{H_D\mathbf{\hat{x}}_{\mathbf{S}_{n-1}}}+\mathbf{z},
\end{equation}
where $\mathbf{z}=\mathbf{e}+\mathbf{H_DS_d}(\mathbf{x_d}-\mathbf{\hat{x}}_{\mathbf{d}_{n-1}})$ is the combined noise for the estimation of $\mathbf{j_I'}$ and $\mathbf{h_D}$ in the $n$-th iteration, and $\mathbf{z}\sim\mathcal{N}(\mathbf{0}_{N\times 1},\frac{1}{N}\sigma_{z}^2\mathbf{I}_{N\times N})$, with \begin{equation}
\label{power of z 1}
\begin{split}
\sigma_{z}^2&=\mathbb{E}[\Vert\mathbf{z}\Vert^2]\\
&=\sigma_e^2+(1-\lambda_S)E_{h_S}d(n-1).
\end{split}
\end{equation}
Then, the parameter vector $\bm{\theta_1} $, which contains all the unknown parameters to be estimated, is defined as
\begin{equation}
\label{theta 1}
\bm{\theta_1}=\begin{bmatrix}
\mathbf{j}_{\mathbf{I_{re}}}'^T,\mathbf{j}_{\mathbf{I_{im}}}'^T,
\mathbf{h}_{\mathbf{D_{re}}}^T,\mathbf{h}_{\mathbf{D_{im}}}^T
\end{bmatrix}^T,
\end{equation}
where 
\begin{equation}
\begin{split}
\label{unknown parameters of theta 1}
\mathbf{j_{I_{re}}'}&=\Re\{\mathbf{j_{I}'}\},\mathbf{j_{I_{im}}'}=\Im\{\mathbf{j_{I}'}\},\\
\mathbf{h_{D_{re}}}&=\Re\{\mathbf{h_D}\},\mathbf{h_{D_{im}}}=\Im\{\mathbf{h_D}\}.
\end{split}
\end{equation}
The Fisher information matrix for $ \bm{\theta_1} $ is given as \cite{7937774}
\begin{equation}
\label{Fisher information of theta 1}
\begin{split}			
\Gamma_{\bm{\theta_1}}=-\mathbf{E}\left[\Delta_{\bm{\theta}}^{\bm{\theta}}\ln{p(\mathbf{y}|\bm{\theta_1})}\right],
\end{split}
\end{equation}
where $ \Delta_{\bm{\theta}}^{\bm{\theta}}f\triangleq \frac{\partial{f}}{\partial{\bm{\theta}}}\left[\frac{\partial{f}}{\partial{\bm{\theta}}}\right]^T$ denotes the second order partial derivative of function $ f $ with respect to vector $ \bm{\theta} $. 

Using \eqref{received signal with data detection errors 2}, the conditional possibility density function of $\mathbf{y}$ given $\bm{\theta_1}$ is derived as 
\begin{equation}
\begin{split}
\label{PDF of y with given theta 1}
p(\mathbf{y}|\bm{\theta_1})=\frac{1}{(\pi \sigma_{z}^2)^N}\times
\exp \bigg\{-\frac{1}{\sigma_{z}^2}(\mathbf{y-u})^H(\mathbf{y-u})\bigg\},
\end{split}
\end{equation}  
where $\mathbf{u}$ is defined from \eqref{received signal with data detection errors 2} as 
\begin{equation}
\begin{split}
\label{u 1}
\mathbf{u}&=\mathbf{y-z}\\
&=\mathbf{T_I}\mathbf{S_I}\mathbf{j_I'}
+\mathbf{H_D}\mathbf{\hat{x}}_{\mathbf{S}_{n-1}}.
\end{split}
\end{equation}
By substituting (\ref{u 1}) into (\ref{PDF of y with given theta 1}), $\Gamma_{\bm{\theta_1}}$ is calculated as
\begin{equation}
\label{Fisher information of theta 2}
\Gamma_{\bm{\theta_1}}=\frac{2}{\sigma_{z}^2}\mathbf{E}\left[\Re\left[
\Big({\frac{\partial{\mathbf{u}}}{\partial{\bm{\theta_1}}}}\Big)^H
\frac{\partial{\mathbf{u}}}{\partial{\bm{\theta_1}^T}}
\right]\right],
\end{equation}
where
\begin{equation}
\begin{split}
\label{partial_u_theta 1}
\frac{\partial{\mathbf{u}}}{\partial{\bm{\theta_1}^T}}=
\begin{bmatrix}
\frac{\partial{\mathbf{u}}}{\partial\mathbf{j}_{\mathbf{I_{re}}}'^T}&
\frac{\partial{\mathbf{u}}}{\partial\mathbf{j}_{\mathbf{I_{im}}}'^T}&
\frac{\partial{\mathbf{u}}}{\partial\mathbf{h}_{\mathbf{D_{re}}}^T}&
\frac{\partial{\mathbf{u}}}{\partial\mathbf{h}_{\mathbf{D_{im}}}^T}
\end{bmatrix},
\end{split}
\end{equation}
and by using (\ref{u 1}), the partial derivative of $\mathbf{u}$ with respect to $\bm{\theta_1}$ is calculated by parts as
\begin{equation}
\label{partial h_D j_c' 1}
\begin{split}
\frac{\partial{\mathbf{u}}}{\partial\mathbf{j}_{\mathbf{c_{re}}}'^T}&=
\mathbf{T_IS_I},\,
\frac{\partial{\mathbf{u}}}{\partial\mathbf{j}_{\mathbf{c_{im}}}'^T}=
j\mathbf{T_IS_I},\\
\frac{\partial{\mathbf{u}}}{\partial\mathbf{h}_{\mathbf{D_{re}}}^T}&=
\mathrm{diag}\{\mathbf{S_p}\mathbf{x_p}+\mathbf{S_d}\mathbf{\hat{x}}_{\mathbf{d}_{n-1}}\}\mathbf{F_S},\\
\frac{\partial{\mathbf{u}}}{\partial\mathbf{h}_{\mathbf{D_{im}}}^T}&=
j\mathrm{diag}\{\mathbf{S_p}\mathbf{x_p}+\mathbf{S_d}\mathbf{\hat{x}}_{\mathbf{d}_{n-1}}\}\mathbf{F_S}.
\end{split}
\end{equation}
Then, the fisher information matrix of $ \mathbf{j}_{\mathbf{I}}'$, which is a sub-matrix of $\Gamma_{\bm{\theta_1}}$, is calculated as
\begin{equation}
\begin{split}
\label{Fisher information of j_I' 1}
\Gamma_{\mathbf{j_I'}}&=\frac{2N}{\sigma_{z}^2}\mathbb{E}
\left[\Re\left\{
\begin{bmatrix}
\frac{\partial{\mathbf{u}^H}}{\partial\mathbf{j}_{\mathbf{I_{re}}}'}
\frac{\partial{\mathbf{u}}}{\partial\mathbf{j}_{\mathbf{I_{re}}}'^T}&
\frac{\partial{\mathbf{u}^H}}{\partial\mathbf{j}_{\mathbf{I_{re}}}'}
\frac{\partial{\mathbf{u}}}{\partial\mathbf{j}_{\mathbf{I_{im}}}'^T}\\
\frac{\partial{\mathbf{u}^H}}{\partial\mathbf{j}_{\mathbf{I_{im}}}'}
\frac{\partial{\mathbf{u}}}{\partial\mathbf{j}_{\mathbf{I_{re}}}'^T}&
\frac{\partial{\mathbf{u}^H}}{\partial\mathbf{j}_{\mathbf{I_{im}}}'}
\frac{\partial{\mathbf{u}}}{\partial\mathbf{j}_{\mathbf{I_{im}}}'^T}
\end{bmatrix}
\right\}\right]\\
&=
\frac{2N}{\sigma_{z}^2}\begin{bmatrix}
\Re\{\mathbf{C}_{K\times K}\}&-\Im\{\mathbf{C}_{K\times K}\}\\
\Im\{\mathbf{C}_{K\times K}\}&\Re\{\mathbf{C}_{K\times K}\}\\
\end{bmatrix},
\end{split}
\end{equation}
where the designed approximation factor $K$ for the SI phase noise in the data transmission stage is set as $K=M-L_S$ (this is explained in section \ref{Data Transmission Stage} in details).
The lower bound for the estimation MSE of $ \mathbf{j_I'} $ in the $n$-th iteration is given as
\begin{equation}
\label{CRLB of j_I' 1}
\begin{split}
C_{n}=\mathbb{E}[\Vert{\mathbf{j'_I}-\mathbf{\hat{j}}'_{\mathbf{I}_{n}}\Vert}^2]\ge
\mathrm{tr}\big\{\Gamma_{\mathbf{j_I'}}^{-1}\big\}=
\frac{\sigma_{z}^2}{N}\mathrm{tr}\big\{\mathbf{C}^{-1}\big\}
\end{split},
\end{equation}
where
\begin{equation} 
\begin{split}
\label{C 1}
\mathbf{C}&=\mathbb{E}
[(\mathbf{T_IS_I})^H\mathbf{T_IS_I}]\\
&=E_{h_I}E_I\cdot\mathbf{I}_{K\times K}.
\end{split}
\end{equation}
Similarly, the fisher information of $ \mathbf{h}_{\mathbf{D}} $ is given as
\begin{equation}
\begin{split}
\label{Fisher information of h_D 1}
\Gamma_{\mathbf{h_{D}}}&=\frac{2N}{\sigma_{z}^2}\mathbb{E}
\left[\Re\left\{
\begin{bmatrix}
\frac{\partial{\mathbf{u}^H}}{\partial\mathbf{h}_{\mathbf{D_{re}}}}
\frac{\partial{\mathbf{u}}}{\partial\mathbf{h}_{\mathbf{D_{re}}}^T}&
\frac{\partial{\mathbf{u}^H}}{\partial\mathbf{h}_{\mathbf{D_{re}}}}
\frac{\partial{\mathbf{u}}}{\partial\mathbf{h}_{\mathbf{D_{im}}}^T}\\
\frac{\partial{\mathbf{u}^H}}{\partial\mathbf{h}_{\mathbf{D_{im}}}}
\frac{\partial{\mathbf{u}}}{\partial\mathbf{h}_{\mathbf{D_{re}}}^T}&
\frac{\partial{\mathbf{u}^H}}{\partial\mathbf{h}_{\mathbf{D_{im}}}}
\frac{\partial{\mathbf{u}}}{\partial\mathbf{h}_{\mathbf{D_{im}}}^T}
\end{bmatrix}
\right\}\right]\\
&=
\frac{2N}{\sigma_{z}^2}\begin{bmatrix}
\Re\{\mathbf{D}_{L_S\times L_S}\}&-\Im\{\mathbf{D}_{L_S\times L_S}\}\\
\Im\{\mathbf{D}_{L_S\times L_S}\}&\Re\{\mathbf{D}_{L_S\times L_S}\}\\
\end{bmatrix},
\end{split}
\end{equation}	
with the lower bound for the estimation MSE of $ \mathbf{h_D } $  in the $n$-th iteration given as
\begin{equation}
\begin{split}
\label{CRLB of h_D 1}
D_{n}=\mathbb{E}[\Vert{\mathbf{h_D}-\mathbf{\hat{h}}_{\mathbf{D}_{n}}\Vert}^2]\ge
\mathrm{tr}\big\{\Gamma_{\mathbf{h_D}}^{-1}\big\}=
\frac{\sigma_{z}^2}{N}\mathrm{tr}\big\{\mathbf{D}^{-1}\big\},
\end{split}
\end{equation}
where 
\begin{equation}
\begin{split}
\label{D 1}
\mathbf{D}&=\mathbb{E}
[(\mathrm{diag}\{\mathbf{S_p}\mathbf{x_p}+\mathbf{S_d}\mathbf{\hat{x}}_{\mathbf{d}_{n-1}}\}\mathbf{F_S})^H\mathrm{diag}\{\mathbf{S_p}\mathbf{x_p}+\mathbf{S_d}\mathbf{\hat{x}}_{\mathbf{d}_{n-1}}\}\mathbf{F_S}]\\
&=E_S\cdot\mathbf{I}_{L_S\times L_S}.
\end{split}
\end{equation}
By using \eqref{power of z 1}, and substituting \eqref{C 1} into \eqref{CRLB of j_I' 1} and \eqref{D 1} into \eqref{CRLB of h_D 1}, we prove Proposition \ref{lower bound of the parameter estimation errors 1}.

\section{Proof of Proposition \ref{Effective SINR 1}}
\label{Proof of the closed-form of SINR}
Using \eqref{residual data-located signal 2}, the average effective SINR of $\mathbf{r}_n$ is given as \cite{4355336}
\begin{equation}
	\label{effective SINR 1}
	\overline{\gamma}(n)=\frac{\mathbb{E}[\Vert{\mathbf{s}\Vert}^2]}
	{\mathbb{E}[\Vert{\mathbf{r}_{n}-\mathbf{s}\Vert}^2]+\mathbb{E}[\Vert{\mathbf{s}-\mathbf{\hat{s}_{n}}\Vert}^2]}
\end{equation} 
where $\mathbf{\hat{s}_{n}}=\mathrm{diag}\{\mathbf{S}_{\mathbf{d}}^T\mathbf{F_S}\mathbf{\hat{h}}_{\mathbf{D}_{n}}\}\mathbf{x_d}$ is defined as the estimated SoI based on $\mathbf{\hat{h}}_{\mathbf{D}_{n}}$. 
It is noticed from \eqref{effective SINR 1} that the detection of the desired data $\mathbf{x_d}$ from $\mathbf{r}_n$ is interferenced by two factors, i.e., the residual SI $\mathbf{r}_{n}-\mathbf{s}$, and the SoI estimation error $\mathbf{s}-\mathbf{\hat{s}_n}$ introduced by the channel estimation error between  $\mathbf{\hat{h}}_{\mathbf{D}_{n}}$ and $\mathbf{h}_{\mathbf{D}}$.
To compute \eqref{effective SINR 1}, the power of the SoI is calculated based on \eqref{approximate model of SoI channel and SoI phase noise 1} as
\begin{equation}
	\label{power of SoI 1}
	\begin{split}
	\mathbb{E}[\Vert{\mathbf{s}\Vert}^2]=&\mathbb{E}[\Vert{\mathrm{diag}\{\mathbf{S}_{\mathbf{d}}^T\mathbf{F_Sh_D}\}\mathbf{x_d}\Vert}^2]\\
	=&\frac{Q}{N}(1-\lambda_S)E_{h_S}E_S,
	\end{split}
\end{equation}
the power of the residual SI is calculated based on \eqref{residual data-located signal 2} as 
\begin{equation}
	\label{power of residual SI 1}
	\begin{split}
	\mathbb{E}[\Vert{\mathbf{r}_{n}-\mathbf{s}\Vert}^2]=&\mathbb{E}[\Vert{\mathbf{S}_{\mathbf{d}}^T\mathbf{T_I}\mathbf{S_I}(\mathbf{j'_I}-\mathbf{\hat{j}}'_{\mathbf{I}_n})+\mathbf{S}_{\mathbf{d}}^T\mathbf{e}\Vert}^2]\\
	=&\frac{Q}{N}E_{h_I}E_IC_{n-1}+\frac{Q}{N}\sigma_e^2,
	\end{split}
\end{equation}
and the power of the error between $\mathbf{s}$ and $\mathbf{\hat{s}_n}$ is calculated as
\begin{equation}
	\label{SoI channel estimation noise 1}
	\mathbb{E}[\Vert{\mathbf{s}-\mathbf{\hat{s}_{n}}\Vert}^2]=\frac{Q}{N}D_{n}E_S.
\end{equation}
By substituting \eqref{power of SoI 1}, \eqref{power of residual SI 1}, and \eqref{SoI channel estimation noise 1} into \eqref{effective SINR 1}, $\overline{\gamma}(n)$ is first derived as 
\begin{equation}
	\label{Effective SINR 2}
	\overline{\gamma}(n)=\frac{(1-\lambda_S)E_{h_S}E_S}
	{C_{n}E_{h_I}E_I+D_{n}E_S+\sigma_{e}^2},
\end{equation} 
where $E_S$, $E_I$, $E_{h_I}$, $E_{h_S}$, $\lambda_I$ and $\lambda_S$ are defined in \eqref{sigma n 1}.

It can be observed in \eqref{Effective SINR 2} that after SI cancellation, the power of the residual receiver noise, which corresponds to the terms in the denominator of (\ref{Effective SINR 2}), can be divided into two parts: 
the first part $\sigma_e^2$ comes from the parameters approximation error in \eqref{sigma n 1}, and is fixed across different iterations; and the second part $C_{n}E_{h_I}E_I+D_{n}E_S$	
is from the noise introduced by the estimation MSE $C_n$ and $D_n$. The derivation of the effective SINR upper bound is based on the utilization of the lower bounds for $C_n$ and $D_n$ in \eqref{lower bound of the parameter lower bound of the parameter estimation errors 1}. It is observed that to compute $d(n-1)=\Vert{\mathbf{x_d}-\mathbf{\hat{x}_{d_{n-1}}}\Vert}^2$ in \eqref{lower bound of the parameter lower bound of the parameter estimation errors 1}, a total of $Q$ data symbols in the desired data vector $\mathbf{x_d}$ along with their all possible detection results in $\mathbf{\hat{x}_{d_{n-1}}}$ should be considered, which is too complicated.

Since identical power is allocated to each OFDM subcarrier in the SI OFDM symbol and in the transmitted OFDM symbol from Node2, it's valid to consider that the received signal on all OFDM subcarriers statically have the identical data detection performance and the identical effective SINR.
Therefore, we can omit the subcarrier indexes for the data symbols, and analyze the effective SINR of arbitrary subcarrier in the sequel. 
Similar to the definition of $d(n-1)$ in \eqref{lower bound of the parameter lower bound of the parameter estimation errors 1}, we define $\overline{d}(n-1)=|X_S-\hat{X}_S^{n-1}|^2$ as the square data detection error in one OFDM subcarrier,
with $X_S$ being the desired data symbol and $\hat{X}_S^{n-1}$ being the detected $X_S$ in the $(n-1)$-th iteration. By substituting $d(n-1)=Q\overline{d}(n-1)$ into \eqref{lower bound of the parameter lower bound of the parameter estimation errors 1}, we have
\begin{equation}
	\label{lower bound of the parameter lower bound of the parameter estimation errors 2}
	\begin{aligned}
		\frac{1}{Q}C_n\ge&\frac{(M-L_S)(\frac{1}{Q}\sigma_e^2+(1-\lambda_S)E_{h_S}\overline{d}(n-1))}{NE_{hI}E_I},\\
		\frac{1}{Q}D_n\ge&\frac{L_S(\frac{1}{Q}\sigma_e^2+(1-\lambda_S)E_{h_S}\overline{d}(n-1))}{NE_S}.
	\end{aligned}
\end{equation}
By substituting \eqref{lower bound of the parameter lower bound of the parameter estimation errors 2} into \eqref{Effective SINR 2}, 
we derive the upper bound for the effective SINR of $\mathbf{r}_n$ and prove Proposition \ref{Effective SINR 1}.

\section{Proof of Proposition \ref{BER Bound 1}}
\label{Appendix: BER Bound}
By utilizing the two data detection results $S^0_{n-1}$ and $S^1_{n-1}$ for the BPSK modulation defined in \eqref{data detection errors 2}, the BER in the $n$-th iteration, $n>1$, is given as
\begin{equation}
	\label{BER i-th 1}
	\begin{split}
		P_n&=p(S_n^1)\\
		&=p(S_{n-1}^0)p(S_n^1|S_{n-1}^0)+P(S_{n-1}^1)p(S_n^1|S_{n-1}^1)\\
		&=(1-P_{n-1})p(S_n^1|S_{n-1}^0)+P_{n-1}p(S_n^1|S_{n-1}^1),
	\end{split}
\end{equation}
where $P_n$ is the BER in the $n$-th iteration.  
Using \eqref{Rayleigh fading channel BER 1}, the conditional probability $p(S_n^1|S_{n-1}^0)$ and $p(S_n^1|S_{n-1}^1)$ is given as
\begin{equation}
	\label{conditional probability 1}
	p(S_n^1|S_{n-1}^0)=f(\gamma_{S_0}(n)),\,
	p(S_n^1|S_{n-1}^1)=f(\gamma_{S_1}(n)),
\end{equation}
where $\gamma_{S_0}(n)$ and $\gamma_{S_1}(n)$ represent the effective SINR in the $n$-th iteration under the two cases $S_{n-1}^0$ and $S_{n-1}^1$, respectively. By calculating the partial derivation of the BER function $f(x)$ with respect to $x$, it is ready to see that $f(x)$ is monotonically decreasing with respect to $x$. By using the effective SINR upper bound $\gamma_0$ and $\gamma_1$ derived in \eqref{effective SINR 5} for the two cases, we have that $f(\gamma_{S_0}(n))$ and $f(\gamma_{S_1}(n))$ are lower bounded, i.e.,
\begin{equation}
	\label{SINR bound 1}
	f(\gamma_{S_0}(n))\ge f(\gamma_0),\, 
	f(\gamma_{S_1}(n))\ge f(\gamma_1).
\end{equation}
Define $P_b=\lim\limits_{n \to \infty}P_n$. By substituting \eqref{conditional probability 1} into \eqref{BER i-th 1} and 
let $n\to\infty$, we can first obtain
\begin{equation}
	\label{BER 1}
	\begin{split}
		P_{b}&=\lim\limits_{n \to \infty}(1-P_{n-1})f(\gamma_{S_0}(n))+
		\lim\limits_{n \to \infty}P_{n-1}f(\gamma_{S_1}(n))\\
		&=(1-P_b)f(\gamma_{S_0}(\infty))+P_bf(\gamma_{S_1}(\infty)).
	\end{split}
\end{equation}
By substituting \eqref{SINR bound 1} into \eqref{BER 1}, $P_b$ is proved to be lower bounded, i.e.,
\begin{equation}
	\label{BER 2}
	\begin{split}
		P_{b}&\ge (1-P_b)f(\gamma_0)+P_bf(\gamma_1)\\
		&\ge \frac{f(\gamma_0)}{1+f(\gamma_0)-f(\gamma_1)},
	\end{split}
\end{equation}
which completes the proof.

\section{Phase Noise Modeling and Computation of the Residual Phase Noise Power}  
\label{Power of the residual phase noise}
In the simulation of this paper, according to many previous works \cite{6937196}, we simply model the phase noise as Wiener process\cite{7815419}, with $\sigma^2=4\pi \Delta f/N$ being the variance and $\Delta f$ being the relative bandwidth of the phase noise,  which reflects the quality of the oscillator \cite{6937196}.

From \eqref{DFT coefficients of the phase noise} and \eqref{sigma n 1}, we notice that the power of the frequency-domain SI phase noise and SoI phase noise are determined by the phase noise $\theta_R(t)$ in the receiver of Node 1, the phase noise $\theta_I(t)$ in the transmitter of Node 1, and the phase noise $\theta_S(t)$ in the transmitter of Node 2. 

For simplify,  we consider the case that all the oscillators at the transceivers of Nodes 1 and 2 have identical quality, and thus $\theta_R(t)$, $\theta_I(t)$, and $\theta_S(t)$ have the same variance $\sigma_{\theta}^2$. Both the cases with two separate oscillators and one common oscillator at one transceiver are considered in the following analysis.

\subsubsection{Separate Oscillators}
Under the case that two separate oscillators are adopted at one transceiver, $\theta_R(t)$, $\theta_I(t)$, and $\theta_S(t)$ are independent of each other \cite{7145898}.
Using \eqref{DFT coefficients of the phase noise}, the power of the frequency-domain SI phase noise in the $k$-th OFDM subcarrier is calculated as \cite{7815419}
\begin{equation}
	\begin{split}
		\label{power of phase noise in frequency band 1}
		\mathbb{E}[|J_I[k]|^2]
		&=\frac{1}{N^2}\sum_{p=0}^{N-1}\sum_{q=0}^{N-1}
		\mathbb{E}[e^{j(\theta_I(pT_s-t_I)-\theta_I(qT_s-t_I))}]\mathbb{E}[e^{j(\theta_R(pT_s)-\theta_R(qT_s))}]e^{-j\frac{2\pi}{N}k(p-q)}\\
		&=\frac{1}{N^2}\sum_{p=0}^{N-1}\sum_{q=0}^{N-1}
		e^{-\sigma_\theta^2|p-q|}e^{-j\frac{2\pi}{N}k(p-q)}\\
		&=\frac{1}{N}+\frac{2}{N^2}\sum_{n=1}^{N-1}(N-n)
		e^{-n\sigma_\theta^2}\cos\bigg(\frac{2\pi kn}{N}\bigg).
	\end{split}
\end{equation}
By using \eqref{power of phase noise in frequency band 1}, the power of the residual SI phase noise, i.e., $\lambda_I$ defined in \eqref{sigma n 1}, is given as 
\begin{equation}
	\begin{split}
		\label{E-ICI 1}
		\lambda_I&=\sum_{k=K}^{N-1}\mathbb{E}[|J_I[k]|^2]\\
		&=1-\frac{K}{N}+\frac{2}{N^2}\bigg\{\sum_{n=1}^{N-1}(N-n)
		e^{-n\sigma_\theta^2}\sum_{k=K}^{N-1}\cos\bigg(\frac{2\pi kn}{N}\bigg)\bigg\}.\\
	\end{split}
\end{equation}
Since all oscillators are considered to have identical quality, the power of the SoI phase noise in the k-th OFDM subcarrier, i.e., $\mathbb{E}[|J_S[k]|^2]$, is identical to $\mathbb{E}[|J_I[k]|^2]$. Then, by substituting $K=1$ into \eqref{E-ICI 1}, the power of the residual SoI phase noise, i.e., $\lambda_S$ defined in \eqref{sigma n 1}, is given as 
\begin{equation}
	\begin{split}
		\label{E[|J_d[0]|^2] 1}
		\lambda_S&=1-\frac{1}{N}+\frac{2}{N^2}\bigg\{\sum_{n=1}^{N-1}(N-n)
		e^{-n\sigma_\theta^2}\sum_{k=1}^{N-1}\cos\bigg(\frac{2\pi kn}{N}\bigg)\bigg\}\\
		&=\frac{1}{N^2}\Bigg[2\frac{e^{-(N+1)\sigma_\theta^2}-(N+1)e^{-\sigma_\theta^2}+N}
		{\Big(e^{-\sigma_\theta^2}-1\Big)^2}-N\Bigg].
	\end{split}
\end{equation}

\subsubsection{Common Oscillator}
Under the case that one common oscillator is adopted at one transceiver, $\theta_S(t)$ is still independent of $\theta_R(t)$ and $\theta_I(t)$, while  $\theta_R(t)=\theta_I(t)$. Since the relationship between $\theta_S(t)$ and $\theta_R(t)$ in this case is the same as the above separate oscillators case, it follows that $\lambda_S$ have the identical value in both the separate oscillators case and common oscillator case. 

Under the common oscillator case, the power of the frequency SI phase noise in the $k$-th OFDM subcarrier is first computed as
\begin{equation}
	\begin{split}
		\label{power of phase noise in frequency band 2}
		\mathbb{E}[|J_I[k]|^2]
		&=\frac{1}{N^2}\sum_{p=0}^{N-1}\sum_{q=0}^{N-1}
		\mathbb{E}[e^{j(\theta_1^{p,q}+\theta_2^{p,q})}]e^{-j\frac{2\pi}{N}k(p-q)},
	\end{split}
\end{equation}
where $\theta_1^{p,q}=\theta_I(pT_s)-\theta_I(qT_s)$, and $\theta_2^{p,q}=\theta_I(pT_s-t_I)-\theta_I(qT_s-t_I)$. To simplify the analysis, we assume $p\ge q$ without loss of generality. If $t_I>NT_s$, for any $p, q\in \{0,1,2,\cdots,N-1\}$, we have $t_I\ge (p-q)T_s$, and $qT_s-t_I<pT_s-t_I<qT_s<pT_s$. Thus, $\theta_1^{p,q}$ and $\theta_2^{p,q}$ are independent of each other, and thus $\mathbb{E}[|J_I[k]|^2]$ and $\lambda_I$ are also identical to their values in the above separate oscillators case. 

If $t_I< NT_s$, for $p, q\in \{0,1,2,\cdots,N-1\}$, some of them still satisfy $t_I>(p-q)T_s$ and the others satisfy $t_I\leq (p-q)T_s$. When $t_I\leq (p-q)T_s$, we have that $qT_s-t_I<qT_s<pT_s-t_I<pT_s$, which means there is an overlap between $\theta_1^{p,q}$ and $\theta_2^{p,q}$. When $t_I\leq (p-q)T_s$, to capitulate the SI phase noise power with this overlap , we rewrite $\theta_1^{p,q}+\theta_2^{p,q}$ as the sum of three independent item, i.e.,
\begin{equation}
	\label{theta1(t)+theta2(t) 1}
	\theta_1^{p,q}+\theta_2^{p,q}=
	[\theta_I(qT_s)-\theta_I(qT_s-t_I)]+
	2[\theta_I(pT_s-t_I)-\theta_I(qT_s)]+
	[\theta_I(pT_s)-\theta_I(pT_s-t_I)].
\end{equation}
Then, according to the above analysis, we have 
\begin{equation}
	\label{auto-correlation of phase noise 2}
	\mathbb{E}[e^{j(\theta_1^{p,q}+\theta_2^{p,q})}]=
	\begin{cases}
		e^{-\sigma_\theta^2|p-q|},\: &p-q<\alpha_I,\\
		e^{-\sigma_\theta^2(|\alpha_I|+|p-q-\alpha_I|)},\: &p-q\ge \alpha_I,\\
	\end{cases}
\end{equation}
where $\alpha_I=\lfloor t_I/T_s \rfloor$ is the relative SI transmission delay. By substituting \eqref{theta1(t)+theta2(t) 1} into \eqref{power of phase noise in frequency band 2}, the power of the frequency SI phase noise is calculated as
\begin{equation}
	\begin{split}
		\label{power of phase noise in frequency band 3}
		\mathbb{E}[|J_I[k]|^2]
		=&\frac{2}{N^2}\sum_{p=0}^{N-1}\sum_{q>p-\alpha_I}^{N-1}
		e^{-\sigma_\theta^2|p-q|}\cos\bigg(\frac{2\pi k(p-q)}{N}\bigg)-
		\frac{1}{N^2}e^{-\sigma_\theta^2\alpha_I}\cos\bigg(\frac{2\pi k\alpha_I}{N}\bigg)\\&+
		\frac{2}{N^2}\sum_{p=0}^{N-1}\sum_{q\leq p-\alpha_I}^{N-1}
		e^{-\sigma_\theta^2(|\alpha_I|+|p-q-\alpha_I|)}\cos\bigg(\frac{2\pi k(p-q)}{N}\bigg).
	\end{split}
\end{equation}



\ifCLASSOPTIONcaptionsoff
  \newpage
\fi

%
\bibliographystyle{IEEEtran}
\bibliography{citation}

%

\end{document}